\documentclass[a4paper,11pt]{article}
\pdfoutput=1 
\usepackage{jcappub}
\usepackage{graphicx}
\usepackage{amsmath}
\usepackage{amssymb}
\usepackage{amsfonts}
\usepackage{mathrsfs}
\usepackage{lscape}
\usepackage{verbatim}
\usepackage{xcolor}
\usepackage{bm}
\usepackage{tabularx}
\usepackage{mathtools}
\usepackage{ulem}
\usepackage{hyperref}

\hypersetup{
  colorlinks = true,
  linkcolor = red,
  urlcolor  = magenta,
  citecolor = blue,
  anchorcolor = blue
}

\newcommand{\mc}{\mathcal}

\newcommand{\be}{\begin{equation}}
\newcommand{\ee}{\end{equation}}

\newcommand{\dd}{{\text{d}}}

\newcommand{\mpl}{m_{\text{P}}}

\newcommand{\aperp}{a_{\perp}}
\newcommand{\apar}{a_{\parallel}}
\newcommand{\mS}{\mathcal{S}}
\newcommand{\mB}{\mathcal{B}}

\newcommand{\mC}{\mathcal{C}}

\renewcommand{\(}{\left(}
\renewcommand{\)}{\right)}
\renewcommand{\[}{\left[}
\renewcommand{\]}{\right]}

\makeatother

\title{Horndeski in motion}

\author[1]{J. Bayron Orjuela-Quintana,\note{Corresponding author.}}
\author[2]{Jose Beltr\'an Jim\'enez,}

\affiliation[1]{Departamento de F\'isica, Universidad del Valle, Ciudad Universitaria Mel\'endez, 760032, Cali, Colombia}

\affiliation[2]{Departamento de F\'isica Fundamental y IUFFyM, Universidad de Salamanca, E-37008 Salamanca, Spain}

\emailAdd{john.orjuela@correounivalle.edu.co}
\emailAdd{jose.beltran@usal.es}

\date{\today}

\abstract{We study a class of homogeneous but anisotropic cosmologies within the family of shift-symmetric Horndeski theories, where the scalar field features an inhomogeneous profile but it preserves a translational symmetry that is realised as a combination of spatial translations and internal shifts. The spatial gradient of the scalar field introduces a preferred direction, so the resulting cosmologies are of the axisymmetric Bianchi I type. The momentum density of these configurations exhibits a universal evolution and an additional component with non-vanishing momentum density is required to have non-trivial effects. We show the relation of these scenarios with cosmologies of non-comoving components and, in particular, we explain how they  provide a specific realisation of moving dark energy models. Among the class of shift-symmetric Horndeski theories, we  analyse in more detail the case of Kinetic Gravity Braiding with emphasis on its application to moving dark energy models and its effects on large scale dark flows as well as the CMB dipole and quadrupole.}
\begin{document}

\maketitle

\section{Introduction}
\label{Sec: Introduction}

The discovery of the current phase of accelerated expansion of the universe represented a major step towards the puzzling dark sector of our cosmological model conformed by dark matter and dark energy. The standard model attributes the accelerated expansion to the presence of a cosmological constant, which is arguably the most natural explanation, but it has the inconvenient of its unnaturally small value~\cite{Weinberg:1988cp, Martin:2012bt}. Impelled by this naturalness problem and the reasonable possibility that some other agent might be driving the accelerated expansion, many models of dark energy, understood in a broad sense, have been proposed~\cite{Clifton:2011jh, Joyce:2014kja}. Scalar fields have played a starring role in this scenario because of their simplicity, their ubiquitous presence in high energy theories and because they naturally comply with the cosmological principle. The complexity of models involving scalar fields has grown from simple quintessence scenarios to the re-discovered Horndeski theories~\cite{Horndeski:1974wa, Deffayet:2011gz} and their extensions~\cite{Zumalacarregui:2013pma, Gleyzes:2014dya, Langlois:2015cwa}. These theories have also been employed to model the other phase of accelerated expansion in the cosmic history, namely, inflation (see e.g. Refs.~\cite{Armendariz-Picon:1999hyi, Arkani-Hamed:2003juy,Alishahiha:2004eh, Kobayashi:2011nu,Martin:2013tda}). In most cases, the scalar field is assumed to be decoupled or interact very weakly with the rest of components of the universe (e.g., photons, baryons or dark matter) and, under these circumstances, a pertinent question is whether the homogeneity frame of the scalar field necessary coincides with the rest frame of other components. For a generic scalar, spatial gradients are, in general, diluted as the universe expands, so the homogeneity frame of the scalar field naturally converges to the rest frame of the other components. This property has been exploited to formulate the effective field theories of inflation~\cite{Cheung:2007st} and dark energy~\cite{Gubitosi:2012hu} where the background scalar field is assumed to be homogeneous from the onset, as preceptive for having homogeneous and isotropic cosmologies. However, if the scalar field has a shift symmetry, its homogeneity frame might differ from the rest frame of other possible components while still preserving cosmological homogeneity thanks to the shift symmetry, although isotropy would have to be sacrificed. This is not to say that the homogeneity frame of the scalar field must differ from that of the other components, but it is a reasonable possibility. As a matter of fact, in the effective field theory of shift-symmetric scalar fields~\cite{Finelli:2018upr}, the scalar field profile is assumed to be homogeneous, again because the interest is in isotropic cosmologies. Certain shift-symmetric subclasses of Horndeski and beyond Horndeski theories have been used in cosmology~\cite{Armendariz-Picon:1999hyi, Scherrer:2004au,Deffayet:2010qz,DeFelice:2010pv,Germani:2011bc,Barreira:2013jma,Martin-Moruno:2015kaa,Germani:2017pwt} (see Ref.~\cite{BorislavovVasilev:2024loq} for a nice and comprehensive review), but the scalar field is routinely assumed to be homogeneous. Even in studies of anisotropic solutions within Horndeski theories, the scalar field profile is assumed to be homogeneous~\cite{Starobinsky:2019xdp, Galeev:2021xit}. This work will be devoted to the exploration of scenarios where the homogeneity of the scalar field is abandoned and it is allowed to acquire an inhomogeneous profile while preserving a diagonal realisation of translations. A natural physical interpretation of these configurations is in terms of a \textit{motion} of the scalar field since the inhomogeneous piece of the scalar field profile generates a non-vanishing component $T_{0i}$ of its energy-momentum tensor, i.e., it carries momentum density. This motivates to dub these scenarios \textit{moving Horndeski models}. It is crucial to realise that these scenarios only make sense if the universe contains another component besides the scalar field because we can always go to the homogeneity frame of the scalar field, that will correspond to the frame where $T_{0i}=0$, by an appropriate choice of coordinates. However, if there is another component, it is not guaranteed that the same choice of coordinates will simultaneously bring both component to their respective rest frames. This can be understood in terms of the symmetries. If that were possible, there would not be any preferred direction in the universe and, hence, we would have an isotropic configuration. On the other hand, if the other component is not in its rest frame in the homogeneity frame of the scalar field, there would be a preferred direction and the universe would exhibit, at most, only a rotational invariance around that preferred axis. The natural metric to describe these cosmologies is then the axisymmetric Bianchi I metric. We will illustrate these points at length with explicit examples in this work. 

In the first part of this work, we will be concerned with the general framework of moving Horndeski theories and we will show some general features as well as different realisations. After a thorough analysis of some general features of these scenarios, we will move on to a specific application to dark energy models. In general, dark energy must provide a somewhat exotic substance with negative pressure to generate expansion and it is usually parametrised as an effective fluid. A fundamental property of this substance, that is usually assumed, is that it has a weak clustering and it barely interacts with standard and dark matter, although these properties can be challenged and they may even provide promising scenarios for explaining some cosmological tensions. However, if we stick to the standard case where dark energy has remained uncoupled from the rest of components of the universe, a pertinent question to ask, that is closely related to the topic of this work, is why would dark energy rest frame agree with the large scale rest frame of matter or the Cosmic Microwave Background (CMB) provided it has always been decoupled. In other words, if dark energy never talked to the other components of the universe, how comes it knows where to sit to be at rest with respect to them on large scales. This possibility was first explored in Ref.~\cite{Maroto:2005kc} where it was shown that a moving dark energy component would generate a cosmological contribution to the CMB dipole and a relative motion between the standard components of the universe. This framework has been further analysed to show its relevance for the CMB quadrupole, the existence of large scale bulk flows, the generation of anisotropic expansion and even the generation of magnetic fields~\cite{BeltranJimenez:2007rsj, BeltranJimenez:2008rei, Harko:2013wsa, Cembranos:2019plq, Cembranos:2019jlp, VillarrubiaRojo:2020peu}. In those works, the moving dark energy model was parametrised as a perfect fluid without any underlying more fundamental field theory description. In this work, we will overcome this shortcoming and provide a more fundamental field-theoretical description in terms of our moving Horndeski scenarios. Another field-theoretical realisation of moving dark energy within bimetric theories has been constructed in Ref.~\cite{Garcia-Garcia:2016dcw}. In this respect, it would seem natural to resort to a homogeneous vector field $A_\mu(t)$ to generate a moving component in the universe, i.e., a non-vanishing $T_{0i}$ that would be generated by $A_i(t)$. However, it has been observed in e.g. Refs.~\cite{BeltranJimenez:2009lpn, BeltranJimenez:2016afo, BeltranJimenez:2018ymu} that $T_{0i}$ is proportional to the equation of motion of the temporal component for homogeneous configurations and, hence, the cosmological momentum density vanishes on-shell. This property represents an obstruction for the construction of moving dark energy models based on vector fields. In this work, we will seize the opportunity to clarify the origin of this obstruction and how shift-symmetric scalars fields (that can be thought of as the longitudinal mode of a vector field) get around it.

From an observational viewpoint, scenarios with moving components are motivated by the accumulated evidence on potential violations of the cosmological principle that include low multipole alignments and hemispherical asymmetry in the CMB, the existence of preferred directions or dipolar modulations in large scale flows~\cite{Krishnan:2022qbv, Krishnan:2022uar, Ebrahimian:2023svi, Allahyari:2023kfm}, the Hubble diagram~\cite{Krishnan:2021jmh, Zhai:2022zif, McConville:2023xav, Sah:2024csa}, quasars, radio galaxies, or fundamental constants among others (see e.g. Refs.~\cite{Aluri:2022hzs, Perivolaropoulos:2021jda} and reference therein). Many of the observed anomalies that might indicate a violation of the cosmological principle suggest that the universe could exhibit some level of anisotropy with a preferred direction, which is precisely what the moving Horndeski scenarios provide.\footnote{It is important to note, however, that similar explanations for these anomalies can also arise from other anisotropic dark energy models. Examples include phenomenological models~\cite{Koivisto:2007bp, Koivisto:2008ig, Koivisto:2010dr, Akarsu:2021max}, and also based on vector fields~\cite{Koivisto:2008xf, Orjuela-Quintana:2021zoe, Garcia-Serna:2023xfw, Gallego:2024gay}, $p$-form fields~\cite{BeltranAlmeida:2019fou, Orjuela-Quintana:2022jrg}, non-Abelian gauge fields~\cite{Orjuela-Quintana:2020klr, Guarnizo:2020pkj, Gomez:2021jbo},  an anisotropic cosmological constant \cite{Rodrigues:2007ny} and even inhomogeneous scalar fields~\cite{Motoa-Manzano:2020mwe}. Also magnetic fields have been employed for similar purposes \cite{Campanelli:2006vb, Campanelli:2007qn}.} Particularly interesting for this work are the claims of anomalously large bulk flows\footnote{Although there are several independent and complementary probes that suggest unexpectedly high bulk flows, we should note the existence of some controversy (see the related discussion in Sec. IV.A. of Ref.~\cite{Aluri:2022hzs}).}~\cite{Kashlinsky:2008ut, Watkins:2008hf, Kashlinsky:2009dw, Kashlinsky:2012gy, Atrio-Barandela:2014nda, Watkins:2023rll} that may challenge the standard model predictions because moving dark energy models can indeed naturally accommodate a large scale relative motion of dark matter with respect to the CMB~\cite{BeltranJimenez:2007rsj}. 

This work is organised as follows: In Section~\ref{Sec: Horndeski in motion}, we will construct and study the general framework of homogeneous cosmologies supported by an inhomogoneous scalar field described by the Horndeski Lagrangians. We will commence in Sec.~\ref{Sec:HorndeskiBianchi}, where we will briefly review Horndeski theories and we will introduce the inhomogeneous scalar field profile that is compatible with homogeneity realised as a combination of spatial translations and internal shifts. In Section~\ref{Sec:UniversalT}, we will show an important property of these scenarios, namely, that the evolution of the momentum density does not depend on the specific theory under consideration but it is determined by homogeneity. In Section~\ref{Sec:MDEconnection}, we will establish a direct connection with models of moving dark energy and non-comoving fluids. We will proceed in Section \ref{sec:mini-superspace} with a mini-superspace analysis in order to clarify how to properly carry out the dimensional reduction, paying attention to the importance of keeping the shift and discussing under what circumstances it can be trivialised. We will conclude our general analysis of the moving Horndeski theories in Section~\ref{sec:duality} with an exploration of the dual formulation in terms of 2-form fields. After presenting a general analysis, we will focus on an application to moving dark energy models in Section~\ref{Sec:MKGB}. We will particularise our general results to the case of Kinetic Gravity Braiding (KGB) theories in Section~\ref{Sec: Moving KGB} and explore the effects on the CMB dipole and quadrupole in Sections~\ref{Sec:Dipole}~and~\ref{Sec: quadrupole}, respectively. In Section~\ref{Sec:IDE}, we will apply our results to a specific KGB dark energy model. We will finally conclude in Section~\ref{Sec:conclusions} with a general discussion of our results.

\section{Horndeski in motion}
\label{Sec: Horndeski in motion}

We will commence our study by constructing homogeneous but anisotropic cosmological solutions described by an axisymmetric Bianchi I metric within the general class of shift-symmetric Horndeski theories. These solutions are supported by a specific inhomogeneous configuration of the scalar field  that is linear in the spatial Cartesian coordinates which leads to a homogeneous physical configuration owed to the shift symmetry. After showing the general properties of these solutions, we will discuss the relation with non-comoving components and some other general features.

\subsection{Bianchi I cosmologies in shift-symmetric Horndeski}
\label{Sec:HorndeskiBianchi}

Let us then commence by briefly reviewing Horndeski theories constructed as the most general scalar-tensor theory explicitly featuring second order field equations for both the metric and the scalar field~\cite{Horndeski:1974wa,Deffayet:2011gz}. The complete Lagrangian of Horndeski theory, as presented in Ref.~\cite{Kobayashi:2011nu}, is expressed as the sum of the following five terms:
\begin{align}
    \mc{L}_2 &= G_2 (\phi, X), \label{Eq: L2}\\
    \mc{L}_3 &= - G_3 (\phi, X) \square \phi, \\
    \mc{L}_4 &= G_4 (\phi, X) R + G_{4 X} (\phi, X) \left[ \( \square \phi \)^2 - \nabla_\mu \nabla_\nu \phi \nabla^\nu \nabla^\mu \phi \right], \\
    \mc{L}_5 &= G_5 (\phi, X) G^{\mu\nu} \nabla_\mu \nabla_\nu \phi \label{Eq: L5}\\
    &- \frac{1}{6} G_{5 X} (\phi, X) \left[ \( \square \phi \)^3 - 3 \( \square \phi \) \nabla_\mu \nabla_\nu \phi \nabla^\nu \nabla^\mu \phi + 2 \nabla^\mu \nabla_\sigma \phi \nabla^\sigma \nabla_\rho \phi \nabla^\rho \nabla_\mu \phi \right] \nonumber. 
\end{align}
Here, $R$ is the Ricci scalar, $G^{\mu\nu}$ is the Einstein tensor, and $G_i(\phi,X)$ (for $i=2,3,4,5$) are arbitrary functions of the scalar field $\phi$ and its kinetic term $X$, defined as $2X \equiv - \nabla_\mu \phi \nabla^\mu \phi$. Throughout this work we will use the shorthand notation $\frac{\partial G_i}{\partial X} \equiv G_{i X}$ to express partial derivatives. 

We will spare the general properties and field equations of the Horndeski theory that can be found in the literature. In this work, we are interested in a particular class of Horndeski theories where the scalar field features a shift-symmetry, ensuring that the Lagrangian remains invariant under the transformation $\phi\mapsto\phi+c$, for some constant $c$. We will refer to Ref.~\cite{BorislavovVasilev:2024loq} for a status report on these theories and their generalities. The shift symmetry is easily implemented by restricting the coupling functions to only depend on $X$, i.e., $G_{i}=G_{i}(X)$.\footnote{Technically, we could allow some explicit dependence on $\phi$. For instance,  $G_3$ could take the form $G_3=c_1\phi+\tilde{G}_2(X)$ with $c_1$ a constant. However, by integrating by parts the term $c_1\phi$ can be included in $G_2$. On the other hand, $G_5$ could also be of the form $G_5=c_1\phi+\tilde{G}_5(X)$. Again, we can integrate by parts so the term $c_1$ would produce $G^{\mu\nu}\partial_\mu\phi\partial_\nu\phi$. Thus, the shift transformation would generate a boundary term in this case. These peculiarities will not be important for us in this work.} Scalar field theories with a shift symmetry on a vacuum state that breaks Lorentz-boosts capture the effective field theory of the phonons in a superfluid (see e.g. \cite{Son:2002zn,Dubovsky:2005xd,Nicolis:2011cs,Pajer:2018egx}). At lowest order in derivatives, these theories simply correspond to the sector described by $G_2(X)$ so we could refer to that sector as the perfect superfluid sector. On the other hand, viscosity and other imperfection effects arise from operators that are higher order in derivatives. Since this is the case of the sectors governed by $G_3(X)$, $G_4(X)$ and $G_5(X)$, we could refer to these terms as imperfect corrections and, thus, the general scenarios under considerations could appealingly be referred to as imperfect superfluids. This terminology has already been used in Ref.~\cite{Pujolas:2011he} in the context of Kinetic Gravity Braiding dark energy models \cite{Deffayet:2010qz}.

The shift-symmetry guarantees, via Noether's theorem, the existence of an on-shell conserved current, which can be written as:
\be
    J^\mu\equiv\frac{1}{\sqrt{-g}}\frac{\delta\mathcal{S}}{\delta\partial_\mu\phi},
\label{eq:defJ}
\ee
where $g$ is the determinant of the metric $g_{\mu\nu}$, and $\mc{S}$ is the action of the theory. It is immediate to see that the scalar field equation of motion adopts the form of the following conservation law:
\be
    \nabla_\mu J^\mu=0.
\ee
For a homogeneous configuration, this conservation equation reduces to:
\be
    \frac{\dd}{\dd t}\Big(\sqrt{-g} J^0\Big)=0,
\ee
so we have that:
\be
    J^0=\frac{Q}{\sqrt{-g}},
\label{eq:J0evolution}
\ee
with $Q$ the conserved charge associated to the shift symmetry. Thus, we find that the evolution of $J^0$ is universal for the whole class of shift-symmetric Horndeski theories. This universality feature has been widely recognised in the literature in different contexts, see e.g. Refs.~\cite{Deffayet:2010qz, Germani:2017pwt, BorislavovVasilev:2024loq}.

Our interest in the shift-symmetric version of Horndeski's theory is that it permits to construct a class of homogeneous but anisotropic cosmologies supported by the following inhomogeneous background configuration:
\be
    \langle\phi\rangle  \equiv \phi(t)+\vec{\lambda}\cdot\vec{x},
\label{eq:fieldprofile}
\ee
with $\vec{\lambda}$ some constant vector, and $\vec{x}$ denoting spatial Cartesian coordinates. Although this configuration is not invariant under spatial translations, it preserves a diagonal translational invariance that involves the shift symmetry and is realised as $\vec{x}\mapsto\vec{x}+\vec{x}_0$, together with $\phi\mapsto\phi-\vec{\lambda}\cdot\vec{x}_0$. This guarantees that all the physical quantities will in turn be homogeneous. Rotational invariance is however broken, meaning the cosmologies supported by this configuration will be homogeneous but anisotropic. This is apparent by looking at the gradient of the scalar field $\partial_\mu\langle\phi\rangle= (\dot{\phi}(t),\lambda_i)$, which is homogeneous but introduces a preferred direction determined by $\vec{\lambda}$. Since the configuration still respects rotations around $\vec{\lambda}$, the appropriate metric to describe these cosmologies will be the axisymmetric Bianchi I family.\footnote{The configuration in Eq.~\eqref{eq:fieldprofile} could accommodate rotating cosmologies as those analysed in e.g. Ref.~\cite{Nicolis:2022gzh}, but we will not consider those in this work.} Anisotropic solutions for Horndeski's theories were considered in~\cite{Starobinsky:2019xdp,Galeev:2021xit}, but the scalar field was assumed to be homogeneous so it cannot sustain an anisotropic expansion by itself. The presence of the inhomogeneous piece with $\vec{\lambda}$ in our configuration \eqref{eq:fieldprofile} offers an advantage in this sense because it can act as a source for the anisotropic expansion. Let us emphasise however that the inclusion of $\vec{\lambda}$ only makes a difference if there are additional components in the universe, as we will discuss in more detail in Sec. \ref{Sec:MDEconnection}. For now, we will focus on the Horndeski sector, but bearing in mind that there could also be other sectors. If we choose the $z$-axis in the direction of $\vec{\lambda}$, we will have $\vec{\lambda}=\lambda\hat{e}_z$, with $\hat{e}_z$ a unit vector pointing along the $z$-axis, and the metric can be written as:
\be
\label{Eq: Bianchi Metric}
    \dd s^2=-\dd t^2+\aperp^2(t)\big(\dd x^2+\dd y^2\big)+\apar^2(t)\dd z^2,
\ee
with $\aperp$ and $\apar$ the scale factors in the transverse and parallel directions to $\vec{\lambda}$. The corresponding expansion rates are $H_{\perp,\parallel} \equiv \frac{\dot{a}_{\perp,\parallel}}{a_{\perp,\parallel}}$, where an overdot represents differentiation with respect to the cosmic time $t$. We can define the isotropic scale factor as:
\be
    a^3 \equiv \aperp^2\apar,
\ee
with the corresponding isotropic expansion rate:
\be
    H \equiv \frac{\dot{a}}{a}=\frac{2H_\perp+H_\parallel}{3}.
\ee
The anisotropy can be described by means of the quantity:
\be
\label{Eq: physical shear}
    \Sigma \equiv \frac{H_\perp-H_\parallel}{3H},
\ee
which measures the deviation from isotropic expansion.

We can evaluate the temporal component of the conserved current, $J^0$, for the axisymmetric Bianchi I metric in Eq.~\eqref{Eq: Bianchi Metric} and the field configuration in Eq.~\eqref{eq:fieldprofile} to obtain $a^3J^0=Q$, so we have a first integral of motion given by: 
\begin{align}
    \frac{Q}{a^3}=&\;G_{2X}\dot{\phi}+6HXG_{3X}+6H^2(1-\Sigma^2)(G_{4X}+2XG_{4XX})\dot{\phi}\\
    &+2H^3(1+\Sigma)^2(1-\Sigma)X(3G_{5X}+2XG_{5XX})\nonumber\\
    &+\frac{2\lambda^2H(1+\Sigma)}{\apar^2}\left[G_{3X}+(1+\Sigma)H\dot{\phi} \, G_{4XX}+(1+\Sigma)(1-2\Sigma)H^2(G_{5X}+XG_{5XX})\right], \nonumber
\label{eq:J0}
\end{align}
where $2X = (\dot{\phi}^2-\lambda^2\apar^{-2})$. The value of $\lambda$ can be absorbed into a re-scaling of the coordinate $z$, which results in a renormalisation of the longitudinal scale factor $\apar$. This means that we could set the magnitude of $\vec{\lambda}$ to one without loss of generality, but we will keep it explicitly for later convenience. Let us emphasise however that this does not mean that the magnitude of $\vec{\lambda}$ is unphysical because, after absorbing it into a re-scaling of the spatial coordinates, it will reappear in other sectors. This is nothing but saying that there is no anisotropic scale invariance in the system.

Now that we have introduced the main ingredients that characterise the considered scenario, we shall exploit the existence of a conserved current to show a remarkable universal relation.

\subsection{Universality of momentum density evolution}
\label{Sec:UniversalT}

In order to physically characterise the Horndeski cosmologies with the profile in Eq.~\eqref{eq:fieldprofile}, it is useful to look at the components of the energy-momentum tensor and, in particular, at its off-diagonal components. Because of our choice of coordinates, the anisotropic stress component is diagonal, so the only non-trivial off-diagonal terms are those associated to an energy-flux or momentum density $T^0{}_i$. By direct computation, we find that those components can be written as:
\be
    T^0{}_i=-J^0\lambda_i.
\label{eq:T0i}
\ee
Since $J^0$ evolves as given in Eq.~\eqref{eq:J0evolution}, irrespective of the form of the coupling functions $G_i(X)$, we conclude that $T^0{}_i$ evolves in the following universal way within the class of shift-symmetric Horndeski theories:
\be
    T^0{}_i=-\frac{Q}{\sqrt{-g}}\lambda_i,
\label{eq:T0ievolution}
\ee
which is the remarkable relation alluded to earlier.  It is interesting to note that the relation in Eq.~\eqref{eq:T0i}, and the associated universal evolution given by Eq.~\eqref{eq:T0ievolution}, is the scalar analogous in shift-symmetric theories of the general relation in vector field theories where $T^0{}_i$ vanishes on-shell because the following relation generically holds:
\begin{equation}
    \sqrt{-g} \, T^0{}_i=-\frac{\delta{\mathcal{S}}}{\delta A_0} A_i,
\end{equation} 
as noticed in e.g. Refs.~\cite{BeltranJimenez:2009lpn, BeltranJimenez:2016afo, BeltranJimenez:2018ymu}. The connection between the shift-symmetric scalar and vector field theories arises if we interpret the gradient of the scalar field $\partial_\mu\phi$ as some vector field $X_\mu$. For instance, this could occur naturally if the shift symmetry is gauged. In that case, the general result for the vector field theories would read: 
\begin{equation}
    \sqrt{-g} \, T^0{}_i=-\frac{\delta{\mathcal{S}}}{\delta X_0} X_i.
\end{equation}
If $X_\mu$ is a genuine vector field, then $\frac{\delta{\mathcal{S}}}{\delta X_0}$ is the equation of motion of the temporal component and, hence, it vanishes on-shell. For the shift symmetric scalar field, however, $\frac{\delta{\mathcal{S}}}{\delta X_0}$ is the temporal component of the conserved current defined in Eq.~\eqref{eq:defJ}, while $X_i=\partial_i\phi=\lambda_i$. The interesting difference between both cases worth emphasising is that, while the result for vector fields prevents the possibility of having a non-trivial $T^0{}_i$, the shift-symmetric scalar fields do permit having a non-trivial $T^0{}_i$ but whose evolution is theory-independent. 

Although the relation in Eq.~\eqref{eq:T0ievolution} has been derived for the axisymmetric Bianchi I metric, it holds for general homogeneous configurations as can be proven from the off-shell conserved currents associated to diffeomorphisms invariance. Let us see how this comes about. Our starting point will be an action of the form: 
\be
\mS[A_\alpha,g_{\mu\nu}]=\int \dd^4x \, L,
\ee
with $L$ the scalar density Lagrangian (thus including the $\sqrt{-g}$ factor) involving up to second derivatives of the metric $g_{\mu\nu}$ and up to first derivatives of the vector field $A_\mu$. Eventually, by identifying the vector field with the gradient of a scalar we will recover the shift-symmetric class of scalar field theories. Under a diffeomorphism parametrised by $\xi^\mu$, the fields transform as follows:
\begin{align}
    \delta_\xi A_\mu=\,&\xi^\lambda\partial_\lambda A_\mu+\partial_\mu\xi^\lambda A_\lambda=\xi^\lambda\nabla_\lambda A_\mu+\nabla_\mu\xi^\lambda A_\lambda,\\
    \delta_\xi g_{\mu\nu}=\,&\xi^\lambda\partial_\lambda g_{\mu\nu}+2\partial_{(\mu}\xi^\lambda g_{\nu)\lambda}=2\nabla_{(\mu}\xi_{\nu)}.
    \label{eq:Difftransf}
\end{align}
The Lagrangian changes by a total derivative under a diffeomorphism, so the action changes by the following boundary term:
\be
    \delta_\xi\mS=\int\dd^4x \, \partial_\alpha\big(\xi^\alpha L\big).
    \label{eq:deltaxiS}
\ee
The total variation of the action under arbitrary variations of the metric and vector field (not necessarily a diffeomorphism) is given by:
\be
    \delta\mS=\int\dd^4x\left[\frac{\delta \mS}{\delta A_\mu}\delta A_\mu+\frac{\delta \mS}{\delta g_{\mu\nu}}\delta g_{\mu\nu}+\partial_\alpha \mB^\alpha\right],
\ee
where $\mB^\alpha$ is the vector density:
\be
\mB^\alpha=\frac{\partial L}{\partial A_{\mu,\alpha}}\delta A_\mu
+\left[\frac{\partial L}{\partial  g_{\mu\nu,\alpha}}-\partial_\beta \left(\frac{\partial L}{\partial  g_{\mu\nu,\alpha\beta}}\right)\right]\delta g_{\mu\nu}+\frac{\partial L}{\partial  g_{\mu\nu,\alpha\beta}}\delta g_{\mu\nu,\beta}.
\ee
We can now use the form of the fields variation under a diffeomorphism given in Eq.~\eqref{eq:Difftransf}, together with the corresponding variation of the action in Eq.~\eqref{eq:deltaxiS}, to find the relation that follows from diffeomorphism invariance, which is given by:
\begin{align}
    \int\dd^4x\left[
    \frac{\delta \mS}{\delta A_\mu}\nabla^\lambda A_\mu
    -\nabla_\mu\left(\frac{\delta \mS}{\delta A_\mu}A^\lambda\right)
    -2\nabla_\mu\left(\frac{\delta \mS}{\delta g_{\mu\lambda}}\right)\right]\xi_\lambda
    +\int\dd^4x \, \partial_\alpha \mC^\alpha(\xi)=0,
\label{eq:xivariation}
\end{align}
with
\be
    \mC^\alpha=\mB^\alpha+\frac{\delta \mS }{\delta A_\alpha}A_\lambda\xi^\lambda+2\frac{\delta \mS}{\delta g_{\alpha\nu}}\xi_\nu-\xi^\alpha L.
\ee
To derive this expression we have used the property $\nabla_\alpha\mathcal\mC^\alpha=\partial_\alpha\mC^\alpha$ which is valid for any vector density of weight one. We can expand $\mC^\alpha$ in derivatives of the gauge parameter $\xi^\lambda$ to express it as:
\be
    \mC^\alpha=\mC_{\lambda}{}^{\alpha}\xi^\lambda+\mC_\lambda{}^{\alpha\mu}\partial_\mu\xi^\lambda +\mC_\lambda{}^{\alpha(\mu\nu)}\partial_\mu\partial_\nu\xi^\lambda,
\ee
where we have defined the coefficients:
\begin{align}
    \mC_\lambda{}^{\alpha\beta\mu} &= 2\frac{\partial L}{\partial g_{\nu\mu,\alpha\beta}}g_{\lambda\nu}, \\
    \mC_\lambda{}^{\alpha\mu} &= \frac{\partial L}{\partial A_{\mu,\alpha}} A_\lambda +\left[\frac{\partial L}{\partial  g_{\mu\nu,\alpha}}-\partial_\beta\left(\frac{\partial L}{\partial  g_{\mu\nu,\alpha\beta}}\right)\right]g_{\lambda\nu}+\frac{\partial L}{\partial g_{\beta\nu,\alpha\mu}}g_{\beta\nu,\lambda}+2\frac{\partial L}{\partial g_{\mu\nu,\alpha\beta}}g_{\lambda\nu,\beta}, \\
    \mC_\lambda{}^\mu &= \frac{\partial L}{\partial A_{\alpha,\mu}} A_{\alpha,\lambda}+\left[\frac{\partial L}{\partial  g_{\alpha\nu,\mu}}-\partial_\beta\left(\frac{\partial L}{\partial  g_{\alpha\nu,\beta\mu}}\right)\right]g_{\alpha\nu,\lambda}+\frac{\partial L}{\partial g_{\alpha\nu,\mu\beta}}g_{\alpha\nu,\lambda\beta}+\frac{\delta\mS}{\delta A_\mu}A_\lambda \nonumber \\
    &+2\frac{\delta \mS}{\delta g_{\mu\nu}}g_{\lambda\nu}-\delta^\mu{}_ \lambda L.\label{eq:coefficients3}
\end{align}
Since $\mC^\alpha$ contributes a boundary term to the variation in Eq.~\eqref{eq:xivariation}, we can choose the gauge parameter $\xi^\mu$ to identically vanish on the boundary so we recover the usual (first) Bianchi identity:
\be
    \frac{\delta \mS}{\delta A_\mu}\nabla^\lambda A_\mu -\nabla_\mu\left(\frac{\delta \mS}{\delta A_\mu}A^\lambda\right) -2\nabla_\mu\left(\frac{\delta \mS}{\delta g_{\mu\lambda}}\right)\equiv0,
\label{eq:firstBianchi}
\ee
that relates the energy-momentum (or gravitational equations) to the matter field equations. We can, however, obtain more identities from the boundary term because, once we have learnt that the bulk contribution to Eq.~\eqref{eq:xivariation} vanishes identically by virtue of Eq.~\eqref{eq:firstBianchi}, we are left with:
\begin{align}
\delta_\xi \mS=\int\dd^4x \, \partial_\alpha\mC^\alpha = \int \dd^4x &\Big[
\partial_\alpha\mC_{\lambda}{}^{\alpha}\xi^\lambda
+\Big(\mC_\lambda{}^\mu+\partial_\alpha\mC_\lambda{}^{\alpha\mu}\Big)\partial_\mu\xi^\lambda\nonumber\\
&+\Big(\mC_\lambda{}^{(\mu\nu)}+\partial_\alpha\mC_\lambda{}^{\alpha(\mu\nu)}\Big)\partial_\mu\partial_\nu\xi^\lambda+\mC_\lambda{}^{(\alpha\mu\nu)}\partial_\alpha\partial_\mu\partial_\nu\xi^\lambda
\Big]=0.
\label{eq:boundaryidentity}
\end{align}
Since the gauge parameter is arbitrary, the condition in Eq.~\eqref{eq:boundaryidentity} implies that the coefficient of each derivative order must vanish independently, giving four additional sets of identities:
\begin{align}
\partial_\alpha\mC_{\lambda}{}^{\alpha}&\equiv0,\label{eq:secondidentities1}\\
\mC_\lambda{}^\mu+\partial_\alpha\mC_\lambda{}^{\alpha\mu}&\equiv0,\label{eq:secondidentities2}\\
\mC_\lambda{}^{(\mu\nu)}+\partial_\alpha\mC_\lambda{}^{\alpha(\mu\nu)}&\equiv0,\label{eq:secondidentities3}\\
\mC_\lambda{}^{(\alpha\beta\mu)}&\equiv0.
\label{eq:secondidentities4}
\end{align}
We can now readily prove the relation $T^0{}_i=J^0X_i$ for homogeneous cosmologies. If we consider a homogeneous cosmology, so spatial partial derivatives vanish, we have from Eq.~\eqref{eq:secondidentities2} the identity:
\be
\mC_i{}^0+\partial_0\mC_i{}^{00}=0.
\ee
However, $\mC_i{}^{00}$ identically vanishes because the constraint in Eq.~\eqref{eq:secondidentities4} gives $\mC_i{}^{000}=0$, so the identity in Eq.~\eqref{eq:secondidentities3} reduces to $\mC_i{}^{00}=0$. Bearing in mind that all spatial derivatives vanish for homogeneous cosmologies, we finally obtain, from the definition in Eq.~\eqref{eq:coefficients3}, the identity:
\be
\mC_i{}^0=\frac{\delta\mS}{\delta A_0}A_i+2\frac{\delta \mS}{\delta g_{0\nu}}g_{i\nu}=0,
\label{eq:Tuniversal}
\ee
which is the relation we wanted to demonstrate, by recalling that $T^{\mu\nu}\equiv \frac{2}{\sqrt{-g}}\frac{\delta \mS}{\delta g_{\mu\nu}}$. The obtained identity generalises straightforwardly to homogeneous configurations containing several vectors fields where we would only need to add the corresponding summation over the different vector fields (for instance tracing over possible internal group structures) in the first term of Eq.~\eqref{eq:T0iuniversal}. This relation shows that, for a vector field theory that is diffeomorphism invariant and whose action contains up to first order derivatives of the vector field and up to second order derivatives of the metric, the momentum density of homogeneous configurations vanishes on-shell. This includes general vector-tensor theories as those considered in e.g. Refs.~\cite{BeltranJimenez:2009lpn,BeltranJimenez:2014iie,Tasinato:2014eka,BeltranJimenez:2015pnp,Hull:2015uwa,Allys:2015sht,BeltranJimenez:2016rff,BeltranJimenez:2016wxw,BeltranJimenez:2016afo,Heisenberg:2016eld,Allys:2016jaq,Allys:2016kbq,Rodriguez:2017wkg,ErrastiDiez:2019ttn,ErrastiDiez:2019trb,GallegoCadavid:2019zke,Aoki:2021wew}, but also more general vector field theories and provides a proof of the observation already made in e.g.  Refs.~\cite{BeltranJimenez:2009lpn, BeltranJimenez:2016afo, BeltranJimenez:2018ymu} that has remained an intriguing feature. On the other hand, if the vector field that we have considered in our derivation in turn describes the gradient of a shift-symmetric theory for a scalar theory, then $\frac{\delta\mS}{\delta A_0}$ no longer coincides with the scalar field equation of motion, but rather gives the temporal component of the conserved current associated to the shift symmetry. Thus, the relation obtained in Eq. \eqref{eq:Tuniversal} also provides a proof for the relation found in Eq. \eqref{eq:T0i} and shows that such a relation applies to a wider class of theories, like e.g. shift symmetric beyond Horndeski or DHOST theories. The crucial property for the validity of Eq.~\eqref{eq:T0i} can be traced back to the existence of a shift symmetry which is then the crucial feature.

Now that we have fully clarified the origin and generality of the remarkable universal relation for shift-symmetric Horndeski theories (and beyond), let us proceed to illustrate why the considered scenario can be naturally interpreted as a moving Horndeski cosmology. In order to do that, we will show its analogy with moving dark energy scenarios, although we should warn that the moving Horndeski framework is more general and does not need to be tied to moving dark energy models.

\subsection{Connection to moving dark energy}
\label{Sec:MDEconnection}

To establish the link between the shift-symmetric Horndeski cosmologies introduced in the precedent sections and the scenarios with moving dark energy considered in Refs.~\cite{Maroto:2005kc,BeltranJimenez:2008rei}, it will be convenient to briefly review the moving dark energy framework. The starting point is a universe filled with several non-interacting fluids with energy-momentum tensors given by:
\begin{equation}
\label{Eq: T perfect fluid}
    T_{(\alpha)}^{\mu\nu} = \left( \rho_{(\alpha)} + p_{(\alpha)} \right) u_{(\alpha)}^\mu u_{(\alpha)}^\nu + p_{(\alpha)} g^{\mu\nu},
\end{equation}
where $\alpha$ (between parenthesis) stands for the different components, such as photons, neutrinos, baryons, dark matter and dark energy. Here, $\rho_{(\alpha)}$, $p_{(\alpha)}$, and $u^\mu_{(\alpha)}$ denote the density, pressure, and 4-velocity of each fluid, respectively. 

In standard cosmological scenarios, the CMB defines a common large-scale rest frame for all the components in the universe, so the 4-velocity of all the fluids coincide and the differences only appear on small (sub-Hubble) scales as peculiar motions. Obviously, this is motivated by the high level of isotropy in the universe suggested by the CMB, provided the dipole is a purely kinematical effect entirely ascribed to our relative motion. If we relax the assumption of a common rest frame, the different fluids can have distinct velocities. In the following, we will assume these relative velocities are small and work at first order in their magnitudes. At this order, the universe is described by a perturbed Friedman-Lema\^itre-Robertson-Walker (FLRW) metric of the following form \cite{Maroto:2005kc}:
\begin{equation}
\label{Eq: Vector Metric}
    \text{d} s^2 = a^2(\eta) \Big[ -\text{d}\eta^2 - 2S_i(\eta) \text{d}\eta \text{d}x^i + \delta_{ij} \text{d}x^i \text{d}x^j \Big],
\end{equation}
with $S_i(\eta)$ a homogeneous perturbation, which only depends on the conformal time $\eta$. The 4-velocities are then given by $u_{(\alpha)}^\mu =\frac{1}{a} \left(-1, \vec{v}_{(\alpha)}\right)$, with $\vec{v}_{(\alpha)}$ the 3-velocity of each fluid that we also assume to be homogeneous to comply with the homogeneity of the metric. The components of the total energy-momentum tensor read:
\begin{align}
    T^{0}_{(\alpha)0} &= - \sum_\alpha \rho_{(\alpha)}, \\
    T^{0}_{(\alpha)i} &=  \sum_\alpha \left(\rho_{(\alpha)} + p_{(\alpha)} \right)\left(v^{(\alpha)}_i-S_i \right), \\
    T^{i}_{(\alpha)0} &= - \sum_\alpha \left(\rho_{(\alpha)} + p_{(\alpha)} \right) v^i_{(\alpha)}, \\
    T^{i}_{(\alpha)i} &= 3 \sum_\alpha p_{(\alpha)}.
\end{align}
From Einstein equations, $G^\mu{}_\nu = 8\pi G T^\mu{}_\nu$, with $G$ denoting the Newton's constant, we obtain the usual Friedman equations at zeroth order:
\begin{equation}
\label{Eq: Friedman Eqs Moving DE}
     \mc{H}^2 = \frac{8\pi G a^2}{3} \sum_\alpha \rho_{(\alpha)}, \quad  \mc{H}' = -\frac{4\pi G a^2}{3} \sum_\alpha \left( \rho_{(\alpha)} + 3 p_{(\alpha)} \right),
\end{equation}
where $\mc{H} \equiv \frac{a'}{a}$ is the Hubble function in conformal time, and a prime denotes differentiation with respect to $\eta$. At first order, the equation $G^0{}_i = 8\pi G T^0{}_i$ gives the constraint:
\be
    0 = 8\pi G\sum_\alpha \left(\rho_{(\alpha)} + p_{(\alpha)} \right)\Big(S_i - v^{(\alpha)}_i\Big),
\ee
which can be solved for $S_i$ to obtain:
\begin{equation}
    S_i = \frac{\sum_\alpha \left(\rho_{(\alpha)} + p_{(\alpha)} \right)v_i^{(\alpha)}}{\sum_\alpha \left(\rho_{(\alpha)} + p_{(\alpha)} \right)}.
\end{equation}
Since $\left(\rho_{(\alpha)} + p_{(\alpha)}\right)$ plays the role of inertial mass density of the corresponding fluid, the metric perturbation $S_i$ can be interpreted as the velocity of the \textit{cosmic center of mass} (CCM)\footnote{The ``$i0$'' component of Einstein equations arrives to the same expression for $S_i$ after replacing $\mc{H}$ and its derivative using the Friedman equations in Eqs.~\eqref{Eq: Friedman Eqs Moving DE}, so it gives redundant information.} and we can define the CCM rest frame by the condition $S_i = 0$. In this frame, the velocities of all the components are related by:
\begin{equation}
    \sum_\alpha \left(\rho_{(\alpha)} + p_{(\alpha)} \right) \vec{v}_{(\alpha)} = 0, 
\label{eq:velocitiesconstraint}
\end{equation}
that nicely resembles the centre of mass relation of the momenta of a system of particles.

Since we are assuming non-interacting fluids, we have the usual momentum conservation equation, $\nabla_\mu T^\mu_{(\alpha) \nu} = 0$, for each component which, at first order in perturbations, yield: 
\begin{eqnarray}
    \rho_{(\alpha)}' + 3\mc{H}\left(\rho_{(\alpha)} + p_{(\alpha)}\right) = 0,\\
    \quad \frac{\dd}{\dd \eta}\left[a^4 \left(\rho_{(\alpha)}+p_{(\alpha)} \right)\left(v_i^{(\alpha)} - S_i\right)\right]=0.
\end{eqnarray}
The first equation is the usual continuity equation for each fluid, while the second equation is nothing but the corresponding momentum conservation equation that determines the evolution of the velocity of the fluid with respect to the CCM. In the CCM rest frame we have:
\be
    \left(\rho_{(\alpha)} + p_{(\alpha)} \right) v_i^{(\alpha)} \propto a^{-4},
\ee
which means that the constraint in Eq.~\eqref{eq:velocitiesconstraint} in turn represents a constraint for the initial velocities of the fluids or, more precisely, for the velocities at some reference time. Let us pause here for a moment and notice the resemblance of this universal evolution, which does not depend on the fluid equation of state, and the expression in Eq.~\eqref{eq:T0ievolution}. In the shift-symmetric Horndeski scenario, the source of anisotropy is characterized by the vector $\vec{\lambda}$, so we can work at first order in $\vec{\lambda}$ for small anisotropies. At that order, equation \eqref{eq:T0ievolution} simply reduces to: 
\be
    T^0{}_i\simeq -\mc{J}^0\vert_{\Sigma=0}\lambda_i,
\ee
where $\mc{J}^0 \equiv J^0/a$ denotes the temporal component of the current $J^\mu$ in conformal time. The latter expression can be further expressed as:
\be
    a^4T^0{}_i\simeq -Q\lambda_i,
\label{eq:T0iuniversal}
\ee
which shows that the relation in Eq.~\eqref{eq:T0ievolution} can be nicely identified as a consequence of momentum conservation for an equivalent fluid. In this equivalence, we clearly see that $\vec{\lambda}$ is naturally identified with the motion of the effective fluid. As a matter of fact, if we fix $Q$ to be the value of the inertial mass $(\rho+p)$ at some reference time (e.g., today), then $\vec{\lambda}$ describes the velocity of the equivalent fluid at that reference time.  It is then apparent that the cosmologies for the shift-symmetric Horndeski theories perfectly fit within the framework of moving dark energy (and non-comoving fluids more generally) so they provide a very natural realisation. This equivalence motivates referring to these cosmologies as \textit{moving Horndeski}.

It is important to bear in mind that the constraint in Eq.~\eqref{eq:velocitiesconstraint} requires the presence of at least one additional component in order to have a non-trivial effect from moving Horndeski since, otherwise, that constraint would impose $\lambda_i = 0$, and we would be back to the usual isotropic Horndeski cosmologies. This also explains why in the study of anisotropic solutions for Horndeski theories performed in Refs.~\cite{Starobinsky:2019xdp, Galeev:2021xit}, the inclusion of $\lambda$ would not have any effect because no additional components are considered and, hence, the equations would impose $\lambda=0$, i.e., the scalar field must be homogeneous. However, if the Horndeski sector is {\it moving} during some epoch, its effects may persist throughout the universe evolution even if the Horndeski scalar eventually disappears. To understand why this happens, it is convenient to notice that, although the combination $a^4 T^0{}_i$ is constant for any fluid, the evolution of the densities and, hence, the velocities will depend on the equation of state of the fluid. If a given fluid has constant equation of state $w$, its density scales as $\rho\propto a^{-3(1+w)}$ and, consequently, its velocity scales as $v_i \propto a^{3w-1}$. For instance, for radiation ($w=1/3$) and pressure-less matter ($w=0$) we have: 
\begin{equation}
\label{Eq: Velocities mr}
\begin{cases}
    \text{radiation: } & \rho_r \propto a^{-4}  \quad \Rightarrow \quad v_i^r = \text{constant}, \\
    \text{matter: } & \rho_m \propto a^{-3}  \quad \Rightarrow \quad v_i^m \propto a^{-1}. 
\end{cases}
\end{equation}
If a moving Horndeski is present before, for example, decoupling, the primordial plasma will have a relative velocity with respect to the CCM. As baryons and photons decouple, the very presence of the moving Horndeski component will generate a relative motion between baryons and photons because of their different equations of state. A similar mechanism might also operate in the early universe in the inflationary, re-heating or post-inflationary era that could generate such relative motions. In this respect, let us notice that, while the velocities of matter components decay with the expansion (although only as $1/a$), the motion of the radiation component persists with its primordial amplitude since its velocity remains constant. In other words, while the motion of matter converges towards the CCM rest frame as the universe expands, radiation maintains a constant relative motion with respect to the CCM frame. This feature opens new possibilities for anisotropic solutions with potentially interesting phenomenological applications. We will take a first step in these phenomenological applications by applying it to dark energy models and will leave further explorations for e.g. the early universe for future work. Before delving into those applications, we will digress a bit to discuss the mini-superspace formulation of the moving Horndeski scenarios and their dual formulations in terms of 2-form fields. This little detour will also allow us to discuss some subtle points of those formulations.

\subsection{Mini-superspace description}
\label{sec:mini-superspace}

Although we can always resort to the covariant equations to obtain and solve the dynamics, it is instructive to discuss the mini-superspace description of the moving Horndeski scenarios, since it entails some potential subtleties that we will have the occasion to clarify. Furthermore, the dimensionally reduced mini-superspace action simplifies many computations and can add additional insights on the system properties, especially regarding symmetries. It is also the starting point for quantum cosmology approach \`a la Wheeler-de Witt. The mini-superspace approach essentially amounts to applying Palais' principle of symmetric criticality \cite{Palais:1979rca} (see also e.g. Ref.~\cite{Fels:2001rv} for applications to General Relativity), so we can perform a symmetry reduction in the action and obtain the equations of motion from such a reduced action. In our case, the residual symmetry group is homogeneity and axial-symmetry, as imposed by the field profile in Eq.~\eqref{eq:fieldprofile}. This is then the symmetry reduction that should be imposed on the metric which should admit three Killing vectors associated to homogeneity and one Killing vector for rotations around the preferred direction. We can use Cartesian coordinates with the preferred direction along the $z$-axis adapted to these Killing vectors, so that translations are generated by $P_{(i)}=\partial_i$, and rotations around the preferred axis by $J_{(z)}=x\partial_y-y\partial_x$. Employing ADM variables, the appropriate Ansatz for the metric can be parameterised as:
\be
\label{Eq: Bianchi I ADM}
    \dd s^2=\left(-N^2+N_i N^i\right)\dd t^2+2N_i\dd t\dd x^i+a^2\left[e^{2\sigma}\Big(\dd x^2+\dd y^2\Big)+e^{-4\sigma}\dd z^2\right],
\ee
where $N(t)$ and $N_i(t)$ represent the lapse function and shift vector, respectively; $a(t)$ the isotropic scale factor and $\sigma(t)$ the shear, that is related to $\Sigma$ introduced in Eq. \eqref{Eq: physical shear} as $\Sigma=\dot{\sigma}/H$. It is important to keep both the lapse and the shift in order to recover the Hamiltonian and momentum constraints. Of course, both can be trivialised by an appropriate diffeomorphism, so we can set them to $N(t)=1$ and $N_i(t)=0$, but this operation can only be performed at the level of the equations and not at the level of the action, i.e., only after carrying out the corresponding variations. Otherwise, the correct equations of motion cannot be recovered. In our set-up, we can align the shift $\vec{N}$ with $\vec{\lambda}$ so we will set $N_i=N_z(t)\delta_{iz}$. For the sake of completeness, we will add a second component besides the Horndeski scalar field and, for simplicity, we will consider a superfluid described by a shift-symmetric scalar $\chi$ with Lagrangian:
\be
    \mathcal{L}=P(\mathcal{X}),\quad\mathcal{X}=-\frac12 (\partial\chi)^2.
\ee
The appropriate configuration for this field is the analogous of that given in Eq.~\eqref{eq:fieldprofile} in order to describe a moving superfluid, so we will consider the following profile:
\be
    \langle\chi\rangle=\chi(t)+\vec{\mu}\cdot \vec{x},
\ee
with $\vec{\mu}$ a constant vector that, in our frame choice, will have the form $\vec{\mu}=\mu \hat{e}_z$. We can now insert our general Ansatz in the action, so we obtain the mini-superspace action:
\be
    \mS=\mS[N,N_z,a,\sigma,\phi,\chi;\lambda,\mu].
    \label{Eq:minisuperspaceaction}
\ee
The precise form of the mini-superspace action is not especially illuminating for our purposes here so we will not reproduce it and will refer to Appendix \ref{App:MovingHorndeskiMinisuperspace} for its explicit form. We will however discuss its key features of relevance for us. Let us notice that both $\lambda$ and $\mu$ do not play the role of dynamical variables but they are rather parameters in the reduced action. Due to the shift-symmetry of the Horndeski sector and the considered matter sector, the variables $\phi$ and $\chi$ only enter with time derivatives and their equations of motion reduce to:
\be
    \frac{\dd Q_\phi}{\dd t}=0,\qquad \frac{\dd  Q_\chi}{\dd t}=0,
    \label{eq:Qconservationmini}
\ee
where $Q_\phi=\frac{\delta \mS}{\delta\dot{\phi}}$ and $Q_\chi=\frac{\delta \mS}{\delta\dot{\chi}}$. On the other hand, the shift $N_z$ and both scalar fields only enter through the combinations $\dot{\phi}-\lambda N_z$ and $\dot{\chi}-\mu N_z$. This is due to a residual symmetry $\phi\to\phi+\epsilon\lambda t$, $\chi\to \chi+\mu \epsilon t$, $N_z\to N_z+\epsilon$, for an arbitrary constant parameter $\epsilon$, that is inherited from the diffeomorphism-invariance of the full Lagrangian. One can corroborate that the gauge transformation $x^i\to x^i-t\epsilon^i $ leads to 
\be
\delta g_{00}=2\epsilon^iN_i,\qquad \delta g_{0i}=\epsilon_i,\qquad \delta g_{ij}=0,\qquad \delta \phi=t\epsilon^i\lambda_i,\qquad \delta \chi=t\epsilon^i\mu_i,
\label{eq:residualgauge}
\ee
which is the quoted residual symmetry with $\vec{\epsilon}=\epsilon\hat{e}_z$. Let us note that the transformation of $g_{00}$ implies that the lapse does not change ($\delta N=0$) and the generated boundary term from the diffeomorphism transformation (we employ the active realisation where only the fields change) is $\partial_i(t\epsilon^i L)=0$ due to the homogeneity of the configuration. This residual symmetry explains why the mini-superspace action is constructed upon the building blocks $\dot{\phi}-\lambda N_z$ and $\dot{\chi}-\mu N_z$ and,  hence, we can relate the variations with respect to the shift $N_z$ with the variations with respect to $\dot{\phi}$ and $\dot{\chi}$, which are precisely the corresponding conserved charges. Thus, the shift equation of motion can be written as
\be
    0=-\frac{\delta \mS}{\delta N_z}=\lambda Q_\phi+\mu Q_\chi,
    \label{eq:mini-superspaceconstraint}
\ee
which in turn gives a constraint for the values of the conserved charges. This constraint arises from the explained residual gauge symmetry and is nothing but the mini-superspace version of the general relation shown in Sec. \ref{Sec:UniversalT}. Let us emphasise that these expressions are general, i.e., we have not fixed neither the lapse nor the shift. We are now permitted to set the shift $N_z=0$, which amounts to going to the CCM rest frame. The constraint in Eq.~\eqref{eq:mini-superspaceconstraint} illustrates once more the fact already explained, that it is necessary to have at least two components in order to have a physically relevant cosmological scenario with moving components. If one of the two components does not move, say $\mu=0$, the momentum constraint forces the other component to either have a vanishing motion $\lambda=0$ or to evolve with a trivial charge $Q_\phi=0$. This is the reason why it is legitimate to set the shift to zero in the mini-superspace action when there is only one matter sector, and it is imposed to have either vanishing motion or vanishing charge. 

The Bianchi identities associated to diffeomorphisms invariance in the mini-superspace read:
\be
    N\frac{\dd}{\dd t}\left(\frac{\delta \mS}{\delta N}\right)+N_z\frac{\dd}{\dd t}\left(\frac{\delta \mS}{\delta N_z}\right)-\dot{a}\frac{\delta \mS}{\delta a}-\dot{\sigma}\frac{\delta \mS}{\delta \sigma}-\dot{\phi}\frac{\delta \mS}{\delta \phi}-\dot{\chi}\frac{\delta \mS}{\delta \chi}=0,
\ee
and they expresses the off-shell relation between the equations of motion. This identity also expresses the fact that both the Hamiltonian and momentum constraints cannot be obtained from the remaining equations via the Bianchi identities, which is of course the statement that they correspond to constraint equations, so fixing either the lapse or the shift at the level of the action is illegitimate in general. In our case, we have shown that the shift can be fixed in the action if the matter sector has either no motion or vanishing charges. This means that the momentum constraint trivialises and, hence, there is no information loss. This is the reason why neglecting the shift in Ref.~\cite{Takahashi:2019vax} was legitimate, where the sector with trivial charge was considered,\footnote{Technically, the authors of Ref.~\cite{Takahashi:2019vax} consider the dual description in terms of a 2-form field (we will analyse this duality in Sec.~\ref{sec:duality}), but the logic is the same.}  and it is the same reasoning employed in Ref.~\cite{Nicolis:2022gzh} to not consider the shift in the mini-superspace formulation of rotating cosmologies in the presence of a single time-dependent scalar or a solid in the unitary gauge.

\subsection{The 2-form dual formulation}
\label{sec:duality}

The moving Horndeski cosmologies introduced above crucially rely on the shift symmetry for the scalar field and it is well-known that massless 2-form fields admit a dualisation to shift symmetric scalar fields. It is thus opportune to ask what the dual formulation of the configuration in Eq.~\eqref{eq:fieldprofile} in terms of a massless 2-form $B_{\mu\nu}$ might be. The answer is given by the following configuration:
\be
    B_{ij}=\frac13B\epsilon_{ijk}x^k+b(t)\epsilon_{ijk}\lambda^k,
\label{eq:Bconfiguration}
\ee
where $B$ is a constant parameter (whose value could be absorbed into a redefinition of the spatial coordinates), $b(t)$ is some function of time and $\lambda^i$ is a constant vector. It is important to notice that the first term in Eq.~\eqref{eq:Bconfiguration} will not, in general, lead to a homogeneous configuration by itself and this is the reason why the 2-form needs to be massless. In that case, there is a gauge symmetry for the 2-form, $B_{\mu\nu}\to B_{\mu\nu}+2\partial_{[\mu}\theta_{\nu]}$, with $\theta_\nu$ an arbitrary 1-form, that allows to restore homogeneity as a combination of spatial rotations and gauge transformations. As a matter of fact, if $b(t)$ is constant, the configuration is not only homogeneous, but also isotropic because the second term in Eq.~\eqref{eq:Bconfiguration} would be a pure gauge mode (see e.g. Ref.~\cite{Aoki:2022ylc} for more details).\footnote{An analogous set-up has been employed in Ref.~\cite{Barros:2023nzr} for the construction of thick brane scenarios with non-trivially realised symmetries on the brane.}  To see this more clearly, it is convenient to recall that the physical quantity for a massless 2-form is its field strength $H_{\mu\nu\rho} \equiv 3\partial_{[\mu}B_{\nu\rho]}$ which, for the configuration in Eq.~\eqref{eq:Bconfiguration}, reads:
\be
    H_{ijk}=B\epsilon_{ijk},\quad H_{0ij}=\dot{b}\epsilon_{ijk}\lambda^k,
\ee
that clearly shows how the 2-form profile in Eq.~\eqref{eq:Bconfiguration} gives rise to a homogeneous but anisotropic configuration, as advertised. It is interesting to notice that the inhomogeneous piece in Eq.~\eqref{eq:Bconfiguration} leads to an isotropic field strength while the homogeneous term is the one responsible for the anisotropic contribution to the field strength. This is the opposite of what happens in the shift-symmetric scalar description where it is the inhomogeneous piece of the scalar field profile in Eq.~\eqref{eq:fieldprofile} the responsible for the anisotropic contributions. This exchange of the roles of the homogeneous and inhomogeneous pieces in the field profile simply reflects the duality relation between both descriptions. We refer to Ref.~\cite{Aoki:2022ylc} for more extensive discussions on this point.

It remains to show that the configuration in Eq.~\eqref{eq:Bconfiguration} also reproduces the universal relation in Eq.~\eqref{eq:T0ievolution} obtained for the moving Horndeski cosmologies. At this point, it is convenient to clarify that the duality between massless 2-forms and shift-symmetric scalars is well-established for theories described by actions that contain up to first order derivatives of the fields (perhaps modulo integration by parts), while the equivalence for theories featuring higher order derivatives is, in general, not possible~\cite{Aoki:2022ylc}. There are however some cases where the equivalence can be established for more general theories (see e.g., Ref.~\cite{Yoshida:2019dxu} for an explicit construction). We will first show the duality for the class of shift-symmetric theories described by $G_2(X)$.\footnote{These theories describe a super-fluid and our configuration in Eq.~\eqref{eq:fieldprofile} would then describe a moving super-fluid. Then, the formulation in terms of a massless 2-form represents the dual formulation of a moving super-fluid, as explained in e.g. Ref.~\cite{Dubovsky:2005xd}.} Let us then consider a 2-form whose dynamics is governed by the following Lagrangian:
\be
    \mathcal{L}=\tilde{G}_2(Y), \quad \text{with} \quad Y \equiv -\frac{1}{12}H_{\mu\nu\rho}H^{\mu\nu\rho}.
    \label{eq:superfluid2form}
\ee
The energy-momentum tensor of the 2-form is given by:
\be
    T_{\mu\nu}=\frac12\tilde{G}_{2 Y}(Y)H_{\mu\alpha\beta}H_\nu{}^{\alpha\beta}+\tilde{G}_2(Y)g_{\mu\nu},
\ee
while the 2-form field equations are:
\be
    \nabla_\mu\Big[\tilde{G}_{2 Y}(Y)H^{\mu\nu\rho}\Big]=0.
\ee
We can now evaluate these equations for the configuration in Eq.~\eqref{eq:Bconfiguration}. We will again choose $\vec{\lambda}=\lambda \hat{e}_z$ so the field equations reduce to the single equation:
\be
    \lambda\frac{\dd}{\dd t}\left[ \frac{\apar^2}{a^3}\dot{b}\tilde{G}_{2Y}(Y)\right]=0,
\ee
that, for $\lambda\neq0$, gives us:
\be
    \frac{\apar^2}{a^3}\tilde{G}_{2 Y}(Y) \dot{b}=Q_B,
\label{eq:QB}
\ee
with $Q_B$ constant. We can now look at the momentum density whose only non-trivial component reads:
\be
    T^0{}_z = -B\frac{\dot{b}\tilde{G}_{2 Y}(Y)}{\aperp^4}\lambda.
\ee
Upon use of Eq.~\eqref{eq:QB}, we obtain:
\be
    T^0{}_z=-B\frac{Q_B}{a^4}\lambda,
\ee
which satisfactorily recovers the universal relation in Eq.~\eqref{eq:T0ievolution} for the moving 2-form described by Eq.~\eqref{eq:superfluid2form} and in the configuration in Eq.~\eqref{eq:Bconfiguration}. This very simple example corresponds to a superfluid in its two dual formulations and has allowed to illustrate how the results of the precedent sections arise for the massless 2-form theories. It is however alluring to see how these results extend to a less trivial example. Thus, we shall now turn our attention to the more interesting case treated in Ref.~\cite{Yoshida:2019dxu} described by the Lagrangian:
\be
\mathcal{L}=-\frac{1}{12\det \mathcal{G}} \mathcal{G}^{\mu\alpha} \mathcal{G}^{\nu\beta} \mathcal{G}^{\rho\gamma} H_{\mu\nu\rho} H_{\alpha \beta \gamma},
\ee
where we have defined the object:
\be
\mathcal{G}_{\mu\nu}\equiv \alpha g_{\mu\nu}+\beta G_{\mu\nu},
\ee
with $\alpha$ and $\beta$ constant parameters and $\det \mathcal{G}$ the determinant of $\mathcal{G}^\mu{}_\nu$. As shown in Ref.~\cite{Yoshida:2019dxu}, this Lagrangian is dual to the scalar theory:
\be
\mathcal{L}=-\frac{1}{2} \mathcal{G}^{\mu\nu}\partial_\mu\phi\partial_\nu\phi,
\ee
so it provides a 2-form dual of a sub-class of Horndeski theories. We will perform our analysis by resorting to the mini-superspace formulation. For completeness, and in order to add further clarifications on how to properly use this approach, we will supplement our 2-form sector with a super-fluid as in Sec.~\ref{sec:mini-superspace}. The mini-superspace action, including the Einstein-Hilbert term for the gravitational sector and the superfluid sector, reads:
\begin{align}
\mathcal{S}=&\int\dd t \left[3 m_\text{P}^2 \frac{a^3}{N}\left(\dot{\sigma}^2 - \frac{\dot{a}^2}{a^2}\right) + \mathcal{P}(\mathcal{X}) - \frac{B^2 N}{2a^3}\left[\alpha+\frac{3\beta}{N^2}\left(\dot{\sigma}^2-\frac{\dot{a}^2}{a^2}\right)\right]^{-1} \right. \nonumber \\
 &\left.+ \frac{e^{-4\sigma}}{2aN} \left(\lambda\dot{b}-B N_z\right)^2\left\{\alpha - \frac{\beta}{N^2}\left[ 2\ddot{\sigma}+3\dot{\sigma}^2+2\frac{\ddot{a}}{a}+\frac{\dot{a}^2}{a^2}+6\frac{\dot{a}\dot{\sigma}}{a}-2\frac{\dot{N}}{N}\left(\dot{\sigma}+\frac{\dot{a}}{a}\right)\right] \right\}^{-1} \right],
\end{align}
where $m_\text{P}^2 = 1/(8\pi G)$ is the reduced Planck mass. Let us notice again that $B$ enters as a parameter of the reduced action and the only dynamical variable associated to the 2-form is $b(t)$. Furthermore, this variable only enters with derivatives so its equation will have the form:
\be
\frac{\dd}{\dd t}\left(\frac{\partial L}{\partial \dot{b}}\right)=0,
\ee
i.e., we recover once again a conservation law. Similarly, the scalar field $\chi$ only enters with derivatives so its equation of motion provides another conservation equation:
\be
\frac{\dd}{\dd t}\left(\frac{\partial L}{\partial \dot{\chi}}\right)=0.
\ee
On the other hand, the equation for the shift, $N_z$, leads to another constraint. In order to unveil such a constraint, we first notice that: 
\be
\frac{\partial L}{\partial N_z}=-\frac{B}{\lambda}\frac{\partial L}{\partial \dot{b}}-\frac{1}{\mu}\frac{\partial L}{\partial \dot{\chi}},
\ee
where we have used that $\mathcal{X}=\frac{1}{2N^2}(\dot{\chi}-\mu N_z)^2-\frac{\mu^2 e^{4\sigma}}{2a^2}$. This relation is of course the version of the universal identity in Eq.~\eqref{eq:Tuniversal} for the system formed by the 2-form and the super-fluid, and the fact that $N_z(t)$, $b(t)$ and $\chi(t)$ only enter the action through the combinations $\lambda\dot{b}-BN_z$ and $\dot{\chi}-\mu N_z$ is a consequence of the residual gauge symmetry analogous to Eq.~\eqref{eq:residualgauge}. Since the shift $N_z$ only enters algebraically in the action, its equation of motion reduces to:
\be
\frac{B}{\lambda}\frac{\partial L}{\partial \dot{b}}+\frac{1}{\mu}\frac{\partial L}{\partial \dot{\chi}}=0,
\label{eq:shifteq}
\ee
where we recognise the conserved quantities $\frac{\partial L}{\partial \dot{b}}\equiv Q_b$ and 
$\frac{\partial L}{\partial \dot{\chi}}\equiv Q_\chi$, and we recover the relation: 
\be
\frac{BQ_b}{\lambda}+\frac{Q_\chi}{\mu}=0.
\ee
We can also easily address the consistency of setting the shift to zero if we restrict to field configurations with $B=0$. For these configurations, the above constraint imposes the triviality of the charge $Q_\chi$ or, in other words, the super-fluid must be at rest with respect to the 2-form field. Under these circumstances, the equation of the shift, Eq.~\eqref{eq:shifteq}, is identically satisfied and, hence, it is consistent to neglect it from the mini-superspace action.

After our slight detour to explore the mini-superspace description of the moving Horndeski scenarios and their dual formulations, let us return to our main route and continue with the application to moving dark energy scenarios.

\section{Moving KGB dark energy}
\label{Sec:MKGB}

In the preceding Sections, we have explored the cosmological implications of the shift-symmetric scalar Horndeski theory with the inhomogeneous scalar field profile specified in Eq.~\eqref{eq:fieldprofile}, as well as its connection to cosmologies featuring non-comoving fluids. In this Section, we will establish a more direct link between these theoretical constructs and models of moving dark energy, providing a concrete realization within the framework of Horndeski theories. Given the stringent observational constraints on the propagation speed of gravitational waves, derived from both the Hulse-Taylor pulsar~\cite{BeltranJimenez:2015sgd} and the LIGO/Virgo collaborations~\cite{LIGOScientific:2017vwq}, we will focus on the Kinetic Gravity Braiding (KGB) subclass of Horndeski theories \cite{Deffayet:2010qz}, which inherently avoid these constraints (see Appendix~\ref{App: GWs in Horndeski} for further discussion). It is worth mentioning however that GWs can still impose severe stability constraints on KGB dark energy models \cite{Creminelli:2019kjy} at non-linear order in perturbations. 

\subsection{Moving KGB}
\label{Sec: Moving KGB}

In the remainder of this work, we restrict our analysis to the shift-symmetric KGB sub-class of Horndeski theories. The dynamics of this model are governed by the Lagrangian:
\be
\label{Eq: KGB action}
    \mc{L}_{\text{KGB}} = G_2(X)-G_3(X)\Box\phi.
\ee
By varying the corresponding action with respect to the metric $g^{\mu\nu}$, the energy-momentum tensor for the scalar field is obtained as:
\begin{equation}
\label{Eq: KGB Energy Tensor}
    T_{\mu\nu}^{(\phi)} = G_2 g_{\mu\nu} + G_{2X} \nabla_\mu \phi \nabla_\nu \phi -  \nabla_\mu \phi \nabla_\nu G_3 + \nabla_\nu \phi \nabla_\mu G_3 + g_{\mu\nu} \nabla_\alpha \phi \nabla^\alpha G_3.
\end{equation}
The corresponding gravitational field equations are expressed as:
\begin{equation}
\label{Eq: Field Eqs}
    m_\text{P}^2 G_{\mu\nu} = T_{\mu\nu}^{(r)} + T_{\mu\nu}^{(m)} + T_{\mu\nu}^{(\phi)},
\end{equation}
where $T_{\mu\nu}^{(r)}$ and $T_{\mu\nu}^{(m)}$ denote the energy-momentum tensors for radiation and pressure-less matter, respectively. In this context, the KGB field, $\phi$, serves as the dynamical component driving the accelerated cosmic expansion.

In our subsequent analysis, we will work with the field equations directly so we are free to fix the shift and the lapse. We will set the shift to zero or, in other words, we will work in the CCM rest frame, and we will also work in conformal time $\eta$, that amounts to fixing the lapse $N(t)=a(t)$. Thus, we will employ the axisymmetric Bianchi I metric, given by:
\be
\label{Eq: Bianchi I}
    \dd s^2=a(\eta)^2\left\{-\dd \eta^2 + \Big[e^{2\sigma}(\dd x^2+\dd y^2)+e^{-4\sigma}\dd z^2\Big]\right\}.
\ee
The non-zero momentum density is manifested through the off-diagonal $T^{0(\phi)}_{\,\ z}$ component of the scalar field's energy-momentum tensor. For the metric in Eq.~\eqref{Eq: Bianchi I}, this component takes the form:
\begin{equation}
    T^{0(\phi)}_{\ \ z} = - \lambda \frac{\phi'}{a^2} \left( G_{2X} + 3 G_{3X} \frac{\mc{H}\phi'}{a^2} \right) +  G_{3X}\left(\mc{H} - 2\sigma'\right)\frac{\lambda^3 e^{4\sigma}}{a^4}.
\end{equation}
As demonstrated in Sec.~\ref{Sec: Horndeski in motion}, the momentum density is intimately related to the conserved current density associated with the general Horndeski theory. For the KGB theory, the current $J_\mu$ can be defined as:
\begin{equation}
\label{Eq: KGB current}
J_\mu = - G_{2X} \nabla_\mu \phi + G_{3X} \Box \phi \nabla_\mu \phi + G_{3X} \nabla_\mu X.
\end{equation}
The temporal component of this current, that will define the conserved charge of the KGB field, reads:
\begin{equation}
    \mc{J}^0 = \frac{\phi'}{a^2} \left( G_{2X} +  3 G_{3X} \frac{\mc{H}\phi'}{a^2} \right) - G_{3X}\left(\mc{H} - 2\sigma'\right)\frac{\lambda^2 e^{4\sigma}}{a^4},
\end{equation}
where we stress the definition $\mc{J} \equiv J^0/a$, which explicitly confirms that the momentum density can be expressed in terms of the current's temporal component as in Eq.~\eqref{eq:T0i}:
\be
T^{0(\phi)}_{\ \ z} = - \mc{J}^0 \lambda.
\ee
The conservation of the homogeneous current density, $\nabla_\mu J^\mu = 0$,  dictates that the temporal component evolves as $a^4 \mc{J}^0 = Q$, where $Q$ is the charge associated to the shift symmetry. Consequently, the momentum density evolves according to:
\be
T^{0 (\phi)}_{\ \ z} = -\frac{Q}{a^4}\lambda,
\label{eq:T0zKGB}
\ee
highlighting the universal nature of the momentum density in terms of the conserved charge. This universal behaviour, as we will show, is essential for quantifying the contribution of the non-comoving components to the CMB dipole.

\subsection{Contribution to the CMB dipole}
\label{Sec:Dipole}

As discussed in Ref.~\cite{Maroto:2005kc}, a non-comoving dark energy component induces a cosmological dipole contribution to the CMB in addition to the usual kinematic component due to our peculiar motion. In order to see how the moving Horndeski component contributes an extra cosmological term to the CMB dipole, it will be useful to briefly review the computation of the CMB dipole via the Sachs-Wolfe effect. 

The leading order contribution to the dipole will be first order in the velocities (or $\lambda$ for the Horndeski component) so that, in the CCM, we can simply assume an isotropic FLRW metric. The energy $\mc{E}$ of a photon propagating along the direction $n^\mu \equiv (-1, \vec{n})$ with $\vec{n}^2 = 1$, as measured by an observer moving with 4-velocity $u^\mu = \frac{1}{a}(-1, \vec{v})$ is given by: 
\begin{equation}
    \mc{E} \equiv u_\mu p^\mu,
\end{equation}
with the photon's 4-momentum:
\begin{equation}
    p^\mu = E \frac{\text{d}x^\mu}{\text{d} \theta},
\end{equation}
where $E$ parametrizes the photon energy, and $x^\mu (\theta)$ describes the photon’s trajectory as a function of the affine parameter $\theta$. Since the FLRW in conformal time is conformal to Minkowski, the photon trajectory, at first order, is simply $x^\mu = \eta n^\mu$ and its 4-momentum becomes: 
\begin{equation}
    p^\mu = \frac{E}{a^2} n^\mu.    
\end{equation}
We can then compute the photon energy at first order as:
\begin{equation}
    \mc{E} = \frac{E}{a}(1 - \vec{v} \cdot \vec{n}).
\end{equation}
Thus, the photon energy experiences a Doppler-like shift due to the observer's motion. This shift contributes to the CMB anisotropies, particularly affecting the dipole in the temperature distribution $\delta T / T_0$, with $T_0$ the monopole temperature. The anisotropy induced in the CMB can be computed via the Sachs-Wolfe effect:
\begin{equation}
    \frac{\delta T}{T_0} = \frac{a_0 \mc{E}_0 - a_\text{dec} \mc{E}_\text{dec}}{a_\text{dec} \mc{E}_\text{dec}},
\end{equation}
where the subscript ``0'' refers to quantities evaluated at present, and ``dec'' refers to quantities evaluated at the time of decoupling. The dipole can be obtained by expanding the above expression to first order in the velocities:
\begin{equation}
    \frac{(\delta T)_{\text{dipole}}}{T_0} \approx \vec{n} \cdot \big(\vec{v}_\text{dec} - \vec{v}_0\big).
\end{equation}
Here, $\vec{v}_\text{dec}$ represents the velocity of the emitter at decoupling, while $\vec{v}_0$ is the observer's velocity at the present time. Assuming that the intrinsic dipole at the last scattering surface is negligible, we have $\vec{v}_\text{dec} \approx \vec{v}_{r, \text{dec}}$, where $\vec{v}_{r, \text{dec}}$ is the radiation fluid's velocity at decoupling in the CCM rest frame. Since the velocity of radiation remains constant throughout the Universe evolution, we have $\vec{v}_{r, \text{dec}}\simeq\vec{v}_{r, \text{0}}$. Furthermore, if the observer is at rest relative to the matter fluid today, we have $\vec{v}_0 \approx \vec{v}_{m, 0}$. Thus, the dipole contribution reduces to: 
\begin{equation}
    \frac{(\delta T)_{\text{dipole}}}{T_0} \approx \vec{n} \cdot \big(\vec{v}_{r, \text{0}} - \vec{v}_{m, 0}\big),
    \label{eq:dipole1}
\end{equation}
which can be interpreted as the relative motion of matter with respect to the CMB at present. Let us notice that, although we have obtained the expression in Eq.~\eqref{eq:dipole1} in the CCM, it remains valid for any frame (at first order) because it is expressed as the difference of the matter and radiation velocities. As explained above, the matter velocity decreases as $1/a$ with the expansion of the universe, so matter components tend to converge towards the CCM,\footnote{This is nothing but the equivalent of why galaxies converge towards the Hubble flow in the standard case. In the moving dark energy scenarios, the convergence occurs towards the CCM frame instead of the Hubble flow.} while radiation maintains a constant velocity relative to the CCM. Consequently, the relative velocity of radiation with respect to all the matter components approximately coincides with its relative CCM velocity, the difference being order $a_{\text{dec}}/a_0\sim 10^{-3}$ with $a_{\text{dec}}$ representing the decoupling time of the corresponding matter sector, and the dipole contribution can be expressed as:
\begin{equation}
    \frac{(\delta T)_{\text{dipole}}}{T_0} \approx \vec{n} \cdot \vec{v}_{r, \text{0}},
\end{equation}
where now the velocity is relative to the CCM.

As discussed in Sec.~\ref{Sec: Horndeski in motion}, the presence of an extra component before decoupling is crucial for the possibility of having a motion of the primordial plasma before decoupling to restore momentum conservation. In our case, the extra source of momentum that makes it possible to have a moving scenario is provided by the KGB field that will contribute to the constraint equation Eq.~\eqref{eq:velocitiesconstraint} with the term \eqref{eq:T0zKGB}. At high redshift, deep within the radiation-dominated epoch, we can identify the radiation velocity with that of the primordial plasma and neglect the matter components so the constraint in Eq.~\eqref{eq:velocitiesconstraint} reads:
\be
    -\frac{Q }{a^4} \vec{\lambda} + \frac{4}{3}\rho_r  \vec{v}_r = 0,
\ee
where we have used that $w_r=1/3$. Using now that $\rho_r\propto a^{-4}$, we can finally obtain the relation: 
\be
\label{Eq: Velocity radiation}
    \vec{v}_r = \frac{3Q}{4 \rho_{r, 0}} \vec{\lambda} = \frac{3\bar{Q}}{4\Omega_{r, 0}} \vec{\lambda},
\ee
where $\bar{Q} \equiv Q/\rho_\text{crit}$, with $\rho_\text{crit}$ denoting the critical density today. This expression relates the (constant) velocity of radiation with the conserved charge of the KGB field and the constant vector $\vec{\lambda}$. Equipped with the relation between the radiation velocity and the moving KGB quantities, we can express the dipole contribution as:
\begin{equation}
    \frac{(\delta T)_{\text{dipole}}}{T_0} \approx \frac{3\bar{Q}}{4\Omega_{r, 0}}\vec{n} \cdot \vec{\lambda},
    \label{eq:dipoleQlambda}
\end{equation}
that gives a direct link between the moving KGB dark energy and the CMB dipole. CMB temperature measurements indicate that $\Omega_{r, 0} \simeq 9 \times 10^{-5}$~\cite{Fixsen:2009}, while the observed amplitude of the CMB dipole is $(\delta T)_{\text{dip}} / T_0 \simeq 1.23 \times 10^{-3}$~\cite{Planck:2018nkj}. This is the maximum value for the dipole that can be generated by the moving KGB field so the model parameters are constrained to be:
\begin{equation}
    \lambda \bar{Q} \lesssim 1.476 \times 10^{-7}.
    \label{eq:dipolebound}
\end{equation}
This constraint saturates if the moving KGB accounts for all of the CMB dipole while peculiar motions only contribute at sub-leading order. If we ascribe the bulk flow excess reported in {\it CosmicFlows4} \cite{Watkins:2023rll}, that amounts to $\sim 380$km/s on spheres of radius $R\sim 200h^{-1}$Mpc, to the moving KGB, we obtain that the radiation velocity relative to the CCM is $v_r\simeq 10^{-3}$ and the bound on the moving KGB parameters is $\lambda \bar{Q}\sim 10^{-7}$, similar to the upper bound in Eq.~\eqref{eq:dipolebound} derived from the CMB dipole. It is important to notice that the value of the conserved charge $\bar{Q}$ is determined by requiring a sufficiently accelerated expansion if the KGB field is to play the role of dark energy, so the obtained constraints in turn translate into an upper bound on $\lambda$. Furthermore, we have not specified any particular KGB model so the resulting constraints on $\lambda \bar{Q}$ are completely general. This is an important consequence of the universal relation in Eq.~\eqref{eq:T0ievolution}.

\begin{figure}[t!]
\centering    
\includegraphics[width=0.6\textwidth]{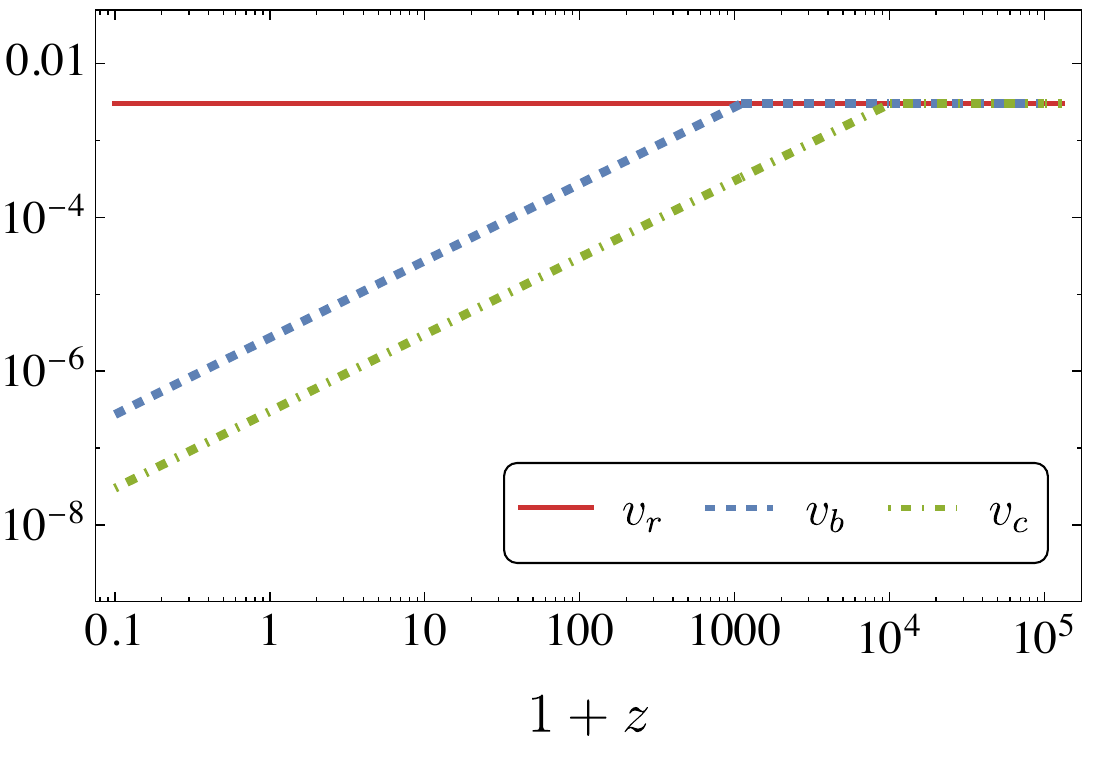}
\caption{Velocity profiles of radiation (red), baryons (blue), and dark matter (green) are shown with $v_r = 1.23 \times 10^{-3}$ in units where $c = 1$. Radiation velocity remains constant, while baryon and dark matter velocities decay as $1/a$ after their decoupling at $z = 1100$ and $z = 10^4$, respectively, approaching the CCM rest frame. Note, however, that dark matter is expected to have decoupled from the primordial plasma at a much higher redshift; the choice of $z = 10^4$ is made for illustrative purposes only.}
\label{Fig: Velocities}
\end{figure}

In the standard model, the observed CMB dipole is fully attributed to our motion relative to the CMB rest frame, defined as the frame with a vanishing dipole~\cite{Planck:2019evm}. In the framework presented here, however, the dipole may include a contribution arising from the difference between our velocity and that of the CMB with respect to the CCM rest frame. To see the magnitude of this difference, in  Fig.~\ref{Fig: Velocities}, we illustrate the velocity evolution of the radiation fluid (red line), baryons (blue line), and dark matter (green line) assuming $v_r = 1.23 \times 10^{-3}$ in units where $c = 1$. The radiation fluid velocity remains constant throughout the cosmic expansion history. Prior to recombination, baryons are tightly coupled to the radiation fluid and they share a common rest frame. However, baryons decouple at $z\simeq 1100$ and, from that moment on, their velocity decays as $1/a$, consistent with Eq.~\eqref{Eq: Velocities mr}. Dark matter is also expected to be coupled to the primordial plasma, but at a much higher redshift,\footnote{Dark matter likely separated from the early plasma before neutrino decoupling ($z = 10^{9}$) but after the electro-weak phase transition ($z = 10^{15}$)~\cite{Bertone:2010zza, ParticleDataGroup:2024cfk}. However, the precise epoch of dark matter decoupling is not relevant for our purposes because its rest frame at present agrees well with the CCM frame, thus making the predictions on the CMB dipole essentially insensitive to the precise dark matter decoupling time; the choice of $z = 10^4$ is made for illustrative purposes. so all these components are expected to share a common rest frame until their respective decouplings. As the velocities of pressure-less matter fluids diminish over time, they asymptotically approach the CCM rest frame.} Notably, the relative velocity difference between these matter fluids and the radiation fluid induces a bulk flow. As commented above, this observation is particularly significant given that numerous studies suggest the presence of an anomalously large bulk flow that could be in tension with the standard $\Lambda$CDM paradigm. Let us emphasise once more that the mere presence of a moving KGB component at decoupling is the trigger to generate relative motions between matter and radiation. The amplitude of the generated bulk flow depends on the relative importance of the moving KGB field at decoupling because it is the amount of KGB momentum density at that time what determines the (constant) amplitude of the radiation velocity. We will illustrate this point with a specific example below, but before, let us study the effects on the quadrupole.

\subsection{Contribution to the CMB quadrupole}
\label{Sec: quadrupole}

At first order in the velocities, only the dipole is affected by the relative motion of the universe components. However, at second order, the quadrupole will also receive a correction due to the relative motions because they generate a certain level of anisotropic expansion at that order. More generally, we expect a contribution to the $\ell-$th multipole of order $\vert v\vert^\ell$ so, if the velocities are small, the effect on the CMB from the relative motions decreases as we go to higher multipoles. In this Section, we will be concerned with the contribution to the quadrupole, where we can still have observable effects, as we will show in the following, while the effects on higher multipoles will be very subdominant with respect to other sources of anisotropy such as e.g. those produced by the primordial inflationary fluctuations.

As previously discussed, non-comoving components can induce anisotropic expansion of the Universe. In the context of this work, such anisotropy is encapsulated by the function $\sigma$ in the axisymmetric Bianchi I metric described in Eq.~\eqref{Eq: Bianchi I}. A key observational consequence of an anisotropic shear is its imprint on the CMB temperature anisotropies, particularly the quadrupole moment, which can be obtained by computing the Sachs-Wolfe effect. If the velocities are small, the leading order contribution to the quadrupole is second order \cite{BeltranJimenez:2007rsj}. Without assuming small velocities, we have shown above that the moving components in the CCM frame give rise to an axisymmetric Bianchi I metric, so we can alternatively compute the Sachs-Wolfe effect for such a metric. The anisotropy induced by the shear of the Bianchi metrics can be expressed as~\cite{Campanelli:2006vb,Pontzen:2007ii,BeltranJimenez:2007rsj,Appleby:2009za}:
\begin{equation}
    \left(\frac{\delta T}{T}\right)_\text{shear} = - \int_{t_\text{dec}}^{t_0} \dd t \, \sigma_{ij} n^i n^j
    \label{eq:deltaTsigma}
\end{equation}
where $t_\text{dec}$ and $t_0$ denote the times of decoupling and today, respectively, and $\sigma_{ij}$ is the shear tensor, defined as:
\begin{equation}
    \sigma_{ij} = \frac{1}{2}\dot{\gamma}_{ij}, \quad \gamma_{ij} = e^{2\beta_i} \delta_{ij}.
\end{equation}
Here, $\gamma_{ij}$ represents the spatial metric of the Bianchi I spacetime, with $\beta_i$ encoding anisotropy along each axis. In our axisymmetric case, $\beta_1 = \beta_2 = \sigma$ and $\beta_3 = -2\sigma$. Consequently, the integrand in the quadrupole expression is determined by $\dot{\sigma}$ and, therefore, the net effect on the photons temperature fluctuation is determined by the variation of the shear between decoupling and today. Furthermore, it is easy to see from Eq.~\eqref{eq:deltaTsigma} that the temperature fluctuation induced by the shear is purely quadrupolar so we can estimate:
\begin{equation}
\label{Eq: CMB quadrupole}
    \frac{(\delta T)_\text{quad}}{T_0} \simeq \sigma_\text{dec} - \sigma_0,
\end{equation}
where the quadrupole is directly tied to the evolution of the geometric shear $\sigma$, which is sourced by the non-comoving components. To determine the evolution equation for $\sigma$, we begin with the trace-free component of the gravitational field equations in Eq.~\eqref{Eq: Field Eqs}. For the left-hand side, the trace-less Einstein tensor component is:
\begin{equation}
    m_\text{P}^2 \left(G^2{}_2 - G^3{}_3 \right) = 3\frac{m_\text{P}^2}{a^2} \left( \sigma'' + 2\mc{H}\sigma' \right).
\end{equation}
For the right-hand side, an anisotropic expansion can accommodate a more general source than a perfect fluid as a source of Einstein equations. In this context, the energy-momentum tensor takes the general form:
\begin{equation}
    T^{(\alpha)}_{\mu\nu} = \left(\rho_{(\alpha)} + p_{(\alpha)} \right) u^{(\alpha)}_\mu u^{(\alpha)}_\nu + p_{(\alpha)} g_{\mu\nu} +2q^{(\alpha)}_{(\mu} u^{(\alpha)}_{\nu)}+ \Pi^{(\alpha)}_{\mu\nu},
\end{equation}
where $q^{(\alpha)}_ \mu$ is the energy flux (or momentum density) and $\Pi^{(\alpha)}_{\mu\nu}$ is the trace-free anisotropic stress tensor. Assuming the absence of intrinsic anisotropic stress contributions from radiation and matter, the relative motion among the different components remains as the sole source for the anisotropic tensor which, in turn, only arises at second order in the velocities \cite{BeltranJimenez:2007rsj}. Thus, radiation and matter source the evolution of $\sigma$ via a term that, in our frame choice where the relative motion takes place along the $z$-direction, can be expressed as:
\begin{equation}
    T^{2}_{(\alpha)2} - T^{3}_{(\alpha)3} = - \left(\rho_{(\alpha)} + p_{(\alpha)} \right) \left(v_z^{(\alpha)} \right)^2 e^{-4\sigma}.
\end{equation}
On the other hand, the anisotropic stress for the scalar field, using Eq.~\eqref{Eq: KGB Energy Tensor}, becomes:
\begin{equation}
    T^{2}_{(\phi)2} - T^{3}_{(\phi)3} = - \frac{\lambda^2 e^{4\sigma}}{a^2} \left[G_{2X} + G_{3X}\left( \frac{\phi''}{a^2} + 2 \frac{\mc{H}\phi'}{a^2} \right) \right].
\end{equation}
Remarkably, by evaluating the spatial component of the density current in Eq.~\eqref{Eq: KGB current}, we can express the anisotropic stress as:\footnote{This expression can be shown to hold generally from the off-shell Bianchi identities obtained in Sec. \ref{Sec:UniversalT} similarly to the universal relation in Eq.~\eqref{eq:T0i}. It is worth clarifying, however, that the quartic and quintic Horndeski Lagrangians do not feature the same general relation of the anisotropic stress and the spatial component of current obtained in Eq.~\eqref{eq:shearJz} for the KGB theories, although we can still use the Bianchi identities to obtain the generalised relations.}
\begin{equation}
    T^{2}_{(\phi)2} - T^{3}_{(\phi)3} = \lambda J^z.
    \label{eq:shearJz}
\end{equation}
However, although this is a general expression valid for any KGB model, it does not exhibit a universality property as the momentum density $T^0{}_i$ in the sense that the evolution of $J^z$ is not constrained by the conservation of $J^\mu$, but it depends on the specific KGB model that is considered.

Combining all contributions to the anisotropic stress, we can obtain the evolution equation for the geometrical shear $\sigma$ as:
\begin{equation}
\label{Eq: Shear Evo}
    \sigma'' + 2\mc{H} \sigma' = S,
\end{equation}
where we have defined the source term:
\be
S=\frac{a^2}{3m_\text{P}^2} \lambda J^z - \sum_\alpha (1 + w_{(\alpha)}) \frac{a^2 \rho_{(\alpha)}}{3 m_\text{P}^2} v^2_{(\alpha)} e^{-4\sigma},
\label{eq:sigmasource}
\ee
where the subscript $z$ on $v_{(\alpha)}$ has been dropped, and $w_{(\alpha)} \equiv p_{(\alpha)} / \rho_{(\alpha)}$ denotes the equation of state of the corresponding fluid. Unlike for the dipole contribution, to fully evaluate the quadrupole contribution we need to specify a KGB model. The reason is that the dipole depends on the value of the conserved charge $Q$ that universally arises thanks to the relation in Eq.~\eqref{eq:T0ievolution}. Moreover, the momentum density also features a universal evolution that allows to eventually express the dipole contribution in terms of the constants $\bar{Q}$ and $\lambda$ as in Eq.~\eqref{eq:dipoleQlambda}. The quadrupole instead depends on $J^z$, whose evolution depends on the specific theory. We can however make some estimates because the shear must be small so we can solve the shear equation in a sort of Born approximation as:
\be
\sigma\simeq\int\frac{\dd\eta}{a^2}\int a^2S\big\vert_{\sigma=0}\dd\eta,
\ee
i.e., by setting the shear to zero in the source term. In this approximation, we can see that the leading order contribution is second order in the velocities and $\lambda$, as we had anticipated. Now we can notice that the contribution to the source term in Eq.~\eqref{eq:sigmasource} from the matter components decays with respect to the radiation contribution. By setting $\sigma=0$ in the source terms, the contribution from radiation and matter can be written as:
\begin{equation}
    a^2S_r = - \frac{4}{3} H_0^2\Omega_{r0}  v_r^2, \quad a^2S_m = - a_\text{dec}^2 \frac{H_0^2\Omega_{m0} }{a} v_r^2.
\end{equation}
where we have used that $v_m=v_r\frac{a_{\text{dec}}}{a}$ because the matter velocity decays as $1/a$ and it has the same velocity as radiation before its decoupling at $a_{\text{dec}}$. As shown in Fig. \ref{Fig: Velocities}, the decoupling of baryons and dark matter happens at different times, but this is not relevant for our argument here. We can then express the matter source as:
\be
S_{m}=\frac{3}{4}\frac{a_{\text{dec}}^2}{a}\frac{\Omega_{m, 0}}{\Omega_{r, 0}}S_r.
\ee
At decoupling, we have $S_{m} \simeq 2.3S_r$, where we have considered the baryons-photons decoupling time, so they are of the same order of magnitude. However, as the universe expands there is a suppressing factor $(a_{\text{dec}}^2/a)$ that makes the matter contribution smaller. For instance, today we have $S_m\simeq2\times10^{-3}S_r$ . Since the contribution to the quadrupole is an integrated quantity, the contribution from radiation will give a good estimate to the quadrupole from the non-Horndeski components. Let us notice that we have overestimated the matter contribution because dark matter (which is the dominant matter component) decouples much before recombination so neglecting the matter contribution is even a better approximation than we have obtained. Thus, we can focus on the radiation contribution that can be computed as:
\be
(\Delta \sigma)_r\simeq-\frac{4}{3}H_0^2\Omega_{r, 0}v_r^2 \int^{\eta_\text{dec}}_{\eta_0}\frac{\dd\eta}{a^2}\int_{\eta_i}^\eta
\dd\hat{\eta}\simeq-\frac{8}{3}\frac{\Omega_{r, 0} v_r^2}{a_{\text{dec}}},
\ee
where $\eta_i$ represents some initial condition on $\sigma$ that does not play any relevant role, we have neglected the KGB contribution to the background evolution and we have used that decoupling occurs in the matter dominated epoch, so we have also neglected the contribution from radiation to the background evolution at decoupling. Thus, if we saturate the bound $v_r\sim 10^{-3}$, we obtain $(\Delta \sigma)_r\sim 10^{-7}$. This value is well-below the CMB quadrupole constraint so that the only potential additional constraint on the moving KGB model must necessarily come from the KGB contribution. Although we need to specify a particular KGB model to compute its contribution to the quadrupole, we can obtain a general constraint as follows:
\begin{eqnarray}
(\Delta \sigma)_{\text{KGB}}&\simeq& \frac{\lambda}{3\mpl^2}\int^{\eta_\text{dec}}_{\eta_0}\frac{\dd\eta}{a^2}\int_{\eta_\star}^\eta
\dd\hat{\eta}\sqrt{-g}J^z=\lambda\bar{Q}H_0^2\int^{\eta_\text{dec}}_{\eta_0}\frac{\dd\eta}{a^2}\int_{\eta_\star}^\eta
\dd\hat{\eta}\frac{J^z}{\mc{J}^0}\nonumber\\
&\lesssim& \lambda\bar{Q}H_0^2\left\vert\frac{J^z}{\mc{J}^0}\right\vert_{\text{max}}\int^{\eta_\text{dec}}_{\eta_0}\frac{\dd\eta}{a^2}\int_{\eta_\star}^\eta
\dd\hat{\eta}\simeq2\frac{\lambda \bar{Q}}{a_{\text{dec}}}\left\vert\frac{J^z}{\mc{J}^0}\right\vert_{\max}, \label{Eq: sigma KGB bound}
\end{eqnarray}
where we have used that $\sqrt{-g} \mc{J}^0 =Q=3\mpl^2H_0^2 \bar{Q}$ and $\vert  J^z/\mc{J}^0\vert_{\text{max}}$ denotes the maximum of the absolute value of $J^z/\mc{J}^0$ in the integration region (i.e., from decoupling till today). Now we can use the CMB dipole bound in Eq.~\eqref{eq:dipolebound} that constrains $\lambda\bar{Q}\lesssim 10^{-7}$ to obtain:
\be
(\Delta \sigma)_{\text{KGB}}\lesssim 10^{-4}\left\vert\frac{J^z}{\mc{J}^0}\right\vert_{\max}.
\ee
Since the CMB quadrupole is $\sim 10^{-5}$, the moving KGB model will be compatible with the CMB quadrupole provided $J_z$ always remains much smaller than $\mc{J}^0$. More specifically, if there is an epoch with $J^z\simeq 10^{-1}\mc{J}^0$, the moving KGB can give a non-negligible contribution to the quadrupole, while if $J^z\ll 10^{-1}\mc{J}^0$ through the entire evolution from decoupling, the CMB quadrupole does not impose any additional constraints on the moving KGB model. Notice that, since $J^z$ is proportional to $\lambda$, the smallness of $\lambda$ impacts the smallness of $J^z$ as well. As a matter of fact, for small $\lambda$ we expect to have $J^z/\mc{J}^0\sim \lambda/\phi'$, that is the small quantity dictating that spatial gradients are much smaller than temporal ones. This means that we can further estimate
\be
(\Delta \sigma)_{\text{KGB}}\lesssim 10^{-4}\frac{\vert\lambda\vert}{\vert\phi'\vert_{\min}}.
\label{eq:Qboundphi}
\ee
Of course, we have made a number of assumptions and approximations, so the obtained bounds should be taken as indicative. As mentioned earlier, a more accurate result requires specifying the KGB model. We will undertake this task in the next section, where we will introduce the imperfect dark energy model~\cite{Deffayet:2010qz} to explicitly show our general arguments. Similarly, but in Appendix~\ref{App: GGC}, we will introduce the galileon ghost condensate model~\cite{Peirone:2019aua}, as a KGB realization where the  moving mechanism driven by scalar field leaves only tiny imprints in the anisotropic expansion, and hence in the CMB quadrupole.

Before proceeding to a specific model, let us point out that the contribution to the quadrupole from the moving KGB model is to be added to the quadrupole generated from inflation, which is a stochastic component. Since the anisotropies are small, the total quadrupole can be computed by linearly adding the inflationary and the moving KGB contributions. If the moving KGB contribution is substantially smaller than the inflationary component, it will have a negligible effect. However, if both contributions are equally important, the stochastic nature of the inflationary component will make the observed total quadrupole in our universe to lie between the sum and the difference of both contributions. This was used in some attempts to explain the low quadrupole value of the CMB \cite{Campanelli:2006vb,BeltranJimenez:2007rsj}.

Let us finally mention that the relative motion of matter and radiation produces an additional contribution to the quadrupole which arises as a second order Doppler effect \cite{BeltranJimenez:2007rsj}. Such a term is $\vec{n} \cdot \big(\vec{v}_\text{dec} - \vec{v}_0\big)(\vec{v}_{\text{dec}}\cdot\vec{n})$. The first factor is again the dipole and the second factor is determined by the observer velocity at decoupling that, as we have discussed above, coincides with the radiation velocity so this contribution is, at most, $\mathcal{O}(10^{-6})$ and, thus, can be safely neglected for our purposes.

\subsection{An explicit example: Imperfect Dark Energy}
\label{Sec:IDE}

After deriving some general properties and results for the moving KGB model, we will turn our attention to a particular realisation to illustrate our general results. As a proxy, we will use the imperfect dark energy model discussed in Ref.~\cite{Deffayet:2010qz} described by the Lagrangian:
\be
\label{Eq: Imperfect DE}
\mathcal{L}_\text{KGB} = -X + \frac{X}{\Lambda^3}\Box\phi,
\ee
with $\Lambda$ some mass scale. It was found in Ref.~\cite{Deffayet:2010qz} that the energy density of the scalar field can peak around radiation-matter equality and has a de Sitter attractor so it will serve the purpose to generate perturbative relative motions. In order to show the feasibility of this scenario to generate relative motions, we will work at leading order in $\lambda$, which means that we can analyse the background cosmology in an FLRW universe. We will then consider a FLRW metric in cosmic time in the following. At leading order in $\lambda$, the conserved charge can be written as:
\be
Q\simeq-a^3\dot{\phi}\left(1-\frac{3H\dot{\phi^2}}{\Lambda^3}\right),
\ee
where we have chosen the sign for convenience. From this expression we can obtain the time-derivative of the scalar field in terms of the conserved charge as:
\be
\dot{\phi}\simeq \frac{\Lambda^3}{6H}\left(1\pm\sqrt{1+\frac{12 Q H}{\Lambda^3 a^3}}\right).
\ee
The minus sign drives the field to an attractor with $\dot{\phi}=0$, which is not the interesting case for us. Instead, the branch with the plus sign evolves to a de Sitter attractor with:
\be
\dot{\phi}=\frac{\Lambda^3}{3 H},
\label{eq:dotphiattractor}
\ee
that is independent of $Q$, as it should given its attractor nature. If this attractor is to drive the accelerated expansion today, we need $\dot{\phi}_0\simeq \frac{\Lambda^3}{3H_0}$. In the regime far from the attractor where $\frac{12 Q H}{\Lambda^3 a^3}\gg1$, we have instead:
\be
\dot{\phi}\simeq\left(\frac{Q\Lambda^3}{3a^3 H}\right)^{1/2},
\label{eq:phidotJdom}
\ee
which is the relation we expect at high redshift, well-inside the radiation dominated epoch, and up to some point in the matter dominated epoch. Indeed, when $\frac{3 Q H}{\Lambda^3 a^3}\simeq 1$, the scalar field is expected to transition to the attractor with $\dot{\phi}=\frac{\Lambda^3}{3 H}$ that will eventually lead to a de~Sitter phase. Notice that this is a natural behaviour because the quantity $\frac{Q H}{\Lambda^3 a^3}$ decreases with the expansion so, if it is larger than one in the early universe, it naturally evolves towards values smaller than one at late times, in compliance with the de Sitter solution being an attractor. This is the generic behaviour utilized in Ref.~\cite{Martin-Moruno:2015kaa} to show the existence of de Sitter attractor. What we need to guarantee in our scenario is that the transition occurs at the correct redshift and, certainly, before today so we have a sufficiently long matter dominated epoch followed by the de Sitter solution.

Let us now look at the energy density of the scalar field, which can be written as
\be
\rho_\phi = -G_2 + \dot{\phi} J^0 = \frac12\dot{\phi}^2+\dot{\phi}\frac{Q}{a^3}.
\ee
In the attractor, where $Q\ll a^3\dot{\phi}$, we have $\rho_\phi \simeq \frac12\dot{\phi}^2$. We can then use Eq.~\eqref{eq:dotphiattractor} to obtain the following relation at the de Sitter attractor:
\be
\rho_\phi \simeq \frac12\left(\frac{\Lambda^3}{3 H_0}\right)^2\simeq 3\mpl^2H_0^2,
\ee
for the field to drive today's cosmic acceleration, which simply means that we need the energy density of the scalar field be of the order of the critical density today. This requirement then fixes the scale of $\Lambda$ to be $\Lambda^3\simeq3\sqrt{6}\mpl H_0^2$. In the regime where the imperfect term dominates, i.e., $Q\gg a^3\dot{\phi}$, the temporal gradient of the scalar field is given by Eq.~\eqref{eq:phidotJdom} and we have:
\be
\rho_\phi \simeq\dot{\phi} J^0\simeq \left(\frac{Q^3 \Lambda^3}{3}\right)^{1/2}(a^9 H)^{-1/2}.
\ee
If the universe is dominated by a fluid with equation of state $w_\text{x}$, being the KGB field subdominant, the obtained expression for the energy density shows that the field evolves with the following equation of state:
\be
w_\phi\simeq\frac{1-w_\text{x}}{4}
\ee
so its energy density evolves as:
\be
\rho_\phi \simeq  \left(\frac{Q^{3}\Lambda^3}{3H_0 \sqrt{\Omega_i}}\right)^{1/2}a^{-3(1+w_\phi)}.
\ee
During the radiation and the matter dominated epochs we find $w_\phi=1/6$ and $w_\phi=1/4$ respectively, in agreement with Ref.~\cite{Deffayet:2010qz}. This means that during radiation domination the energy density of the Horndeski field grows with respect to that of radiation until equality time and, from that moment on, its energy density decreases with respect to the dust dominant component. In order to avoid conflicts with pre-recombination (such as Big Bang Nucleosynthesis) and early dark energy bounds, a safe bound is to impose that the KGB field contributes less than $10\%$ to the energy budget at equality, i.e., $\rho_{\text{eq}}\lesssim0.1\rho_{r,\text{eq}}$. This leads to the bound:
\be
\label{Eq: Q estimation}
\left(\frac{Q}{H_0\mpl\sqrt{a_{\text{eq}}}}\right)^{3/2}\lesssim0.1\quad\Rightarrow\quad\frac{Q}{H_0\mpl}\lesssim 2.3\times10^{-4},
\ee
This bound on $Q$ allows us to estimate when the transition $\frac{12 Q H}{\Lambda^3 a^3}\simeq 1$ occurs, i.e., the transition from the large $Q$ regime to the de Sitter attractor. If we saturate the bound on $Q$ we find that the transition happens at redshift around 5, so it is consistent with our picture. Furthermore, the bound on $Q$ together with the dipole bound in Eq.~\eqref{eq:dipolebound} constrains $\lambda$ to be:
\be
\label{Eq: lambda estimation}
\frac{\lambda}{3H_0\mpl}\lesssim 10^{-3}.
\ee
Let us notice that $\dot{\phi}$ decreases through the universe evolution until it reaches its minimum at the attractor, where it remains constant. Since we have $\dot{\phi}_0\simeq H_0\mpl$ at the attractor, the obtained bound on $\lambda$ in fact guarantees that $\lambda\ll\dot{\phi}$ at all times and, hence, the approximation of small $\lambda$ is valid and consistent. We can however consistently saturate the bounds and generate relative motions with $v_r$ up to $10^{-3}$ so we can still have observable effects on the CMB dipole and bulk flows. The contribution to the CMB quadrupole is expected to be $\lesssim 10^{-7}$ according to Eq.~\eqref{eq:Qboundphi} so we do not have further constraints from the quadrupole.

Now, to enhance our understanding of the model's predictions and validate the general estimates presented above, we perform a comprehensive phase-space analysis. For clarity and to maintain a direct focus on the calculation of the CMB quadrupole contribution, we summarize the primary results of the dynamical analysis here, while detailed computations are provided in Appendix~\ref{App: Dynamical System}.

We commence our dynamical analysis by introducing the following dimensionless variables:
\begin{equation}
\label{Eq: Variables}
    x_1 \equiv -\frac{\phi'^2}{6 m_\text{P}^2 \mc{H}^2}, \ \ x_3 \equiv \frac{\phi'^3/\Lambda^3 }{a^2 m_\text{P}^2 \mc{H}}, \ \ \Omega_r \equiv \frac{a^2 \rho_r}{3 m_\text{P}^2 \mc{H}^2}, \ \ \Omega_b \equiv \frac{a^2 \rho_b}{3 m_\text{P}^2 \mc{H}^2}, \ \ u \equiv \frac{\lambda^2 e^{4\sigma}}{\phi'^2}, \ \ \Sigma \equiv \frac{\sigma'}{\mc{H}}. 
\end{equation}
The evolution of the system is governed by the following set of first-order differential equations:
\begin{align}
    \frac{\text{d}x_1}{\text{d}N} &= 2 x_1 (\epsilon_\phi - h_\phi), \label{Eq: x_1 Eq} \\
    \frac{\text{d}x_3}{\text{d}N} &= x_3 (3\epsilon_\phi - h_\phi - 2), \\
    \frac{\text{d}\Omega_r}{\text{d}N} &= -2 \Omega_r (1 + h_\phi), \\
    \frac{\text{d}\Omega_b}{\text{d}N} &= -\Omega_b (1 + 2h_\phi), \\
    \frac{\text{d} u}{\text{d}N} &= 2u (2\Sigma - \epsilon_\phi), \\
    \frac{\text{d} \Sigma}{\text{d}N} &= -\Sigma(2 + h_\phi) - \frac{1}{3}u\left[6x_1 + x_3(2 + \epsilon_\phi)\right] - \sum_{\alpha = b, c, r} \(1 + w_{(\alpha)}\) \Omega_{(\alpha)} v^2_{(\alpha)}, \label{Eq: Sigma Eq}
\end{align}
Here, $N$ denotes the number of $e$-folds, defined via $\text{d}N \equiv \mathcal{H} \text{d}\eta$. To close the system, we introduce auxiliary parameters $\epsilon_\phi \equiv \phi''/(\mc{H} \phi')$ and $h_\phi \equiv \mc{H}'/\mc{H}^2$, which are expressed in terms of the variables as follows:
\begin{align}
    q_s \epsilon_\phi &= 3x_1 \left[8 - \left(3 - 2 u + 3 u^2 \right) x_3 \right] + x_3\left[u^2 x_3 (2 \Sigma - 3) - 3 (1 + x_3 + 3 \Sigma^2 + \Omega_r) \right] \\
    &+ x_3\left[ u\left(1 + x_3 (2 - 6 \Sigma) + \Sigma(19 \Sigma - 16) + \Omega_r \right)\right] \nonumber, \\
    \frac{3}{2} q_s h_\phi &= -9 (u - 3) x_1^2 + 3 x_1 \left[(u - 3) x_3 (u^2 x_3 - 6) + 6 u x_3 \Sigma + 9 \Sigma^2 + 
    3 (1 + \Omega_r)\right] \\ 
    &- x_3^2 \left[-9 + u (3 - 6 \Sigma (1 + 2 \Sigma) + u (1 + \Sigma(-4 + 7 \Sigma) + \Omega_r))\right] \nonumber \\
    &+ x_3 \left[u^2 x_3^2 \left(u - 2 u \Sigma - 3\right) + 9 \left(1 + 3 \Sigma^2 + \Omega_r\right) \right], \nonumber \\
    q_s &\equiv -12 x_1 + x_3 [ x_3(u - 1) (3 + u) - 12].
\end{align}
The effective equation of state, $w_\text{eff}$, and the dark energy equation of state, $w_\text{DE}$, are given by:
\begin{align}
    w_\text{eff} &\equiv \frac{p_\text{T}}{\rho_\text{T}}= - \frac{1 + 2 h_\phi + 3 \Sigma^2}{3\left(1 - \Sigma^2 \right)}, \\
    w_\text{DE} &\equiv \frac{p_\phi}{\rho_\phi} = \frac{3(3x_1 + x_3(1 - \epsilon_\phi)) - u(3x_1 + x_3(1-\epsilon_\phi) - 6\Sigma x_3)}{9(x_1 + x_3) + u(9x_1 - 3x_3 +6\Sigma x_3)}, \label{Eq: wDE}
\end{align}
where $p_\text{T} \equiv \sum_i p_i$, $\rho_\text{T} \equiv \sum_i \rho_i$, with $i=r, c, b, \phi$, and $p_\phi$ is the scalar field pressure [see Eq.~\eqref{Eq: IDE p}]. To illustrate the model's contributions to the CMB quadrupole [Eq.~\eqref{Eq: CMB quadrupole}], we numerically solve the autonomous system. Assuming the absence of primordial anisotropy ($\Sigma_i = 0$), as we expect such anisotropy to emerge at late times due to relative motion among cosmological components, we specify the initial conditions for the remaining variables deep within the radiation-dominated epoch ($z = 1.3 \times 10^{5}$):
\begin{equation}
\label{Eq: Init Cond}
    \Omega_{r, i} = 0.975, \ \ \Omega_{b, i} = 0.004, \ \ x_{1, i} = -1.2 \times 10^{-13}, \ \ x_{3, i} = 3 \times 10^{-3}, \ \ u_i \approx 0.26,
\end{equation}
which yields $\Omega_{c, i} \approx 0.018$. These initial conditions ensure consistency with the observed energy budget today and saturate the CMB dipole constraint in Eq.~\eqref{eq:dipolebound} via the moving mechanism, where $v_r = 1.23 \times 10^{-3}$. To integrate the non-autonomous system, we account for the decoupling history of the components: baryons decouple from radiation at the surface of last scattering ($z = 1100$), while dark matter decouples earlier, around $z = 10^{4}$. After decoupling, both components decay as $1/a$.

\begin{figure*}[t!]
\centering    
{\includegraphics[width=0.47\textwidth]{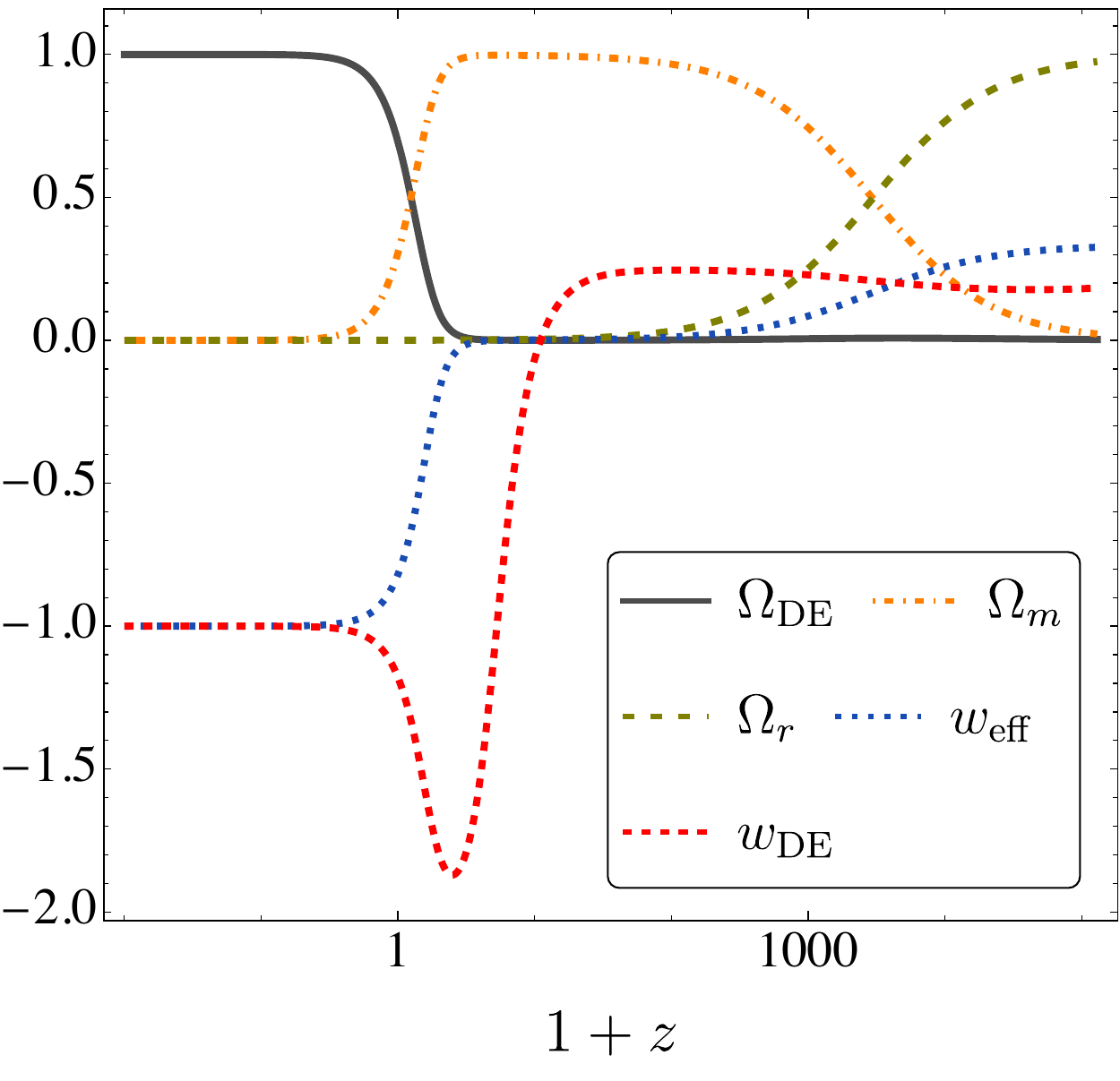}} \hfill
{\includegraphics[width=0.49\textwidth]{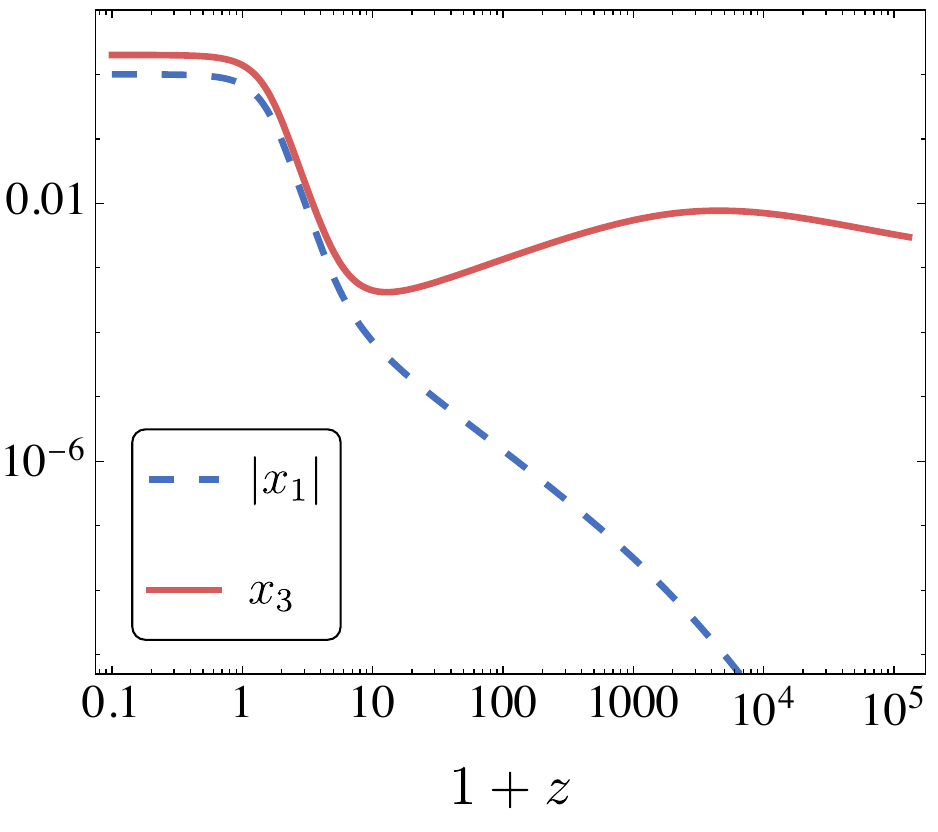}}
\caption{The time evolution of $\Omega_r$, $\Omega_m \equiv \Omega_c + \Omega_b$, $\Omega_\text{DE}$, the equations of state $w_\text{eff}$ and $w_\text{DE}$ (in the left panel), and the variables $|x_1|$, and $x_3$ (in the right panel). At early times, when radiation is the dominant fluid in the cosmic budget ($w_\text{eff} \approx 1/3$), the scalar field evolution is dominated by $x_3$. During the matter-dominated epoch, the scalar field crosses the phantom line approaching $w_\text{DE} = -2$. As time progresses, $|x_1|$ take over the dynamics, and the system transitions to a de Sitter expansion driven by the scalar field, with $w_\text{eff} = w_\text{DE} = -1$.}
\label{Fig: IDE Dynamics}
\end{figure*}

\begin{figure*}[t!]
\centering    
\includegraphics[width=0.475\textwidth]{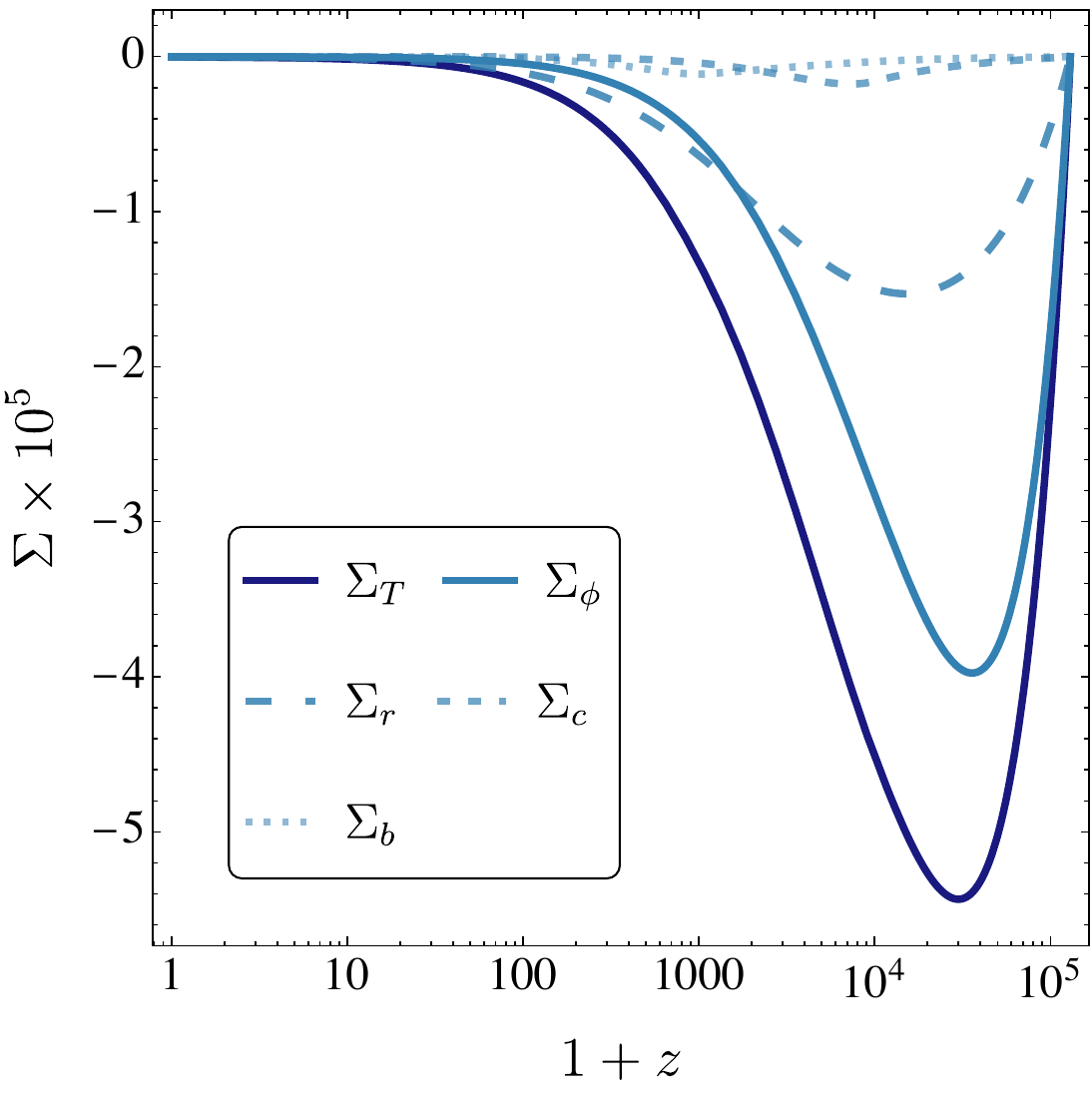} \hfill
{\includegraphics[width=0.475\textwidth]{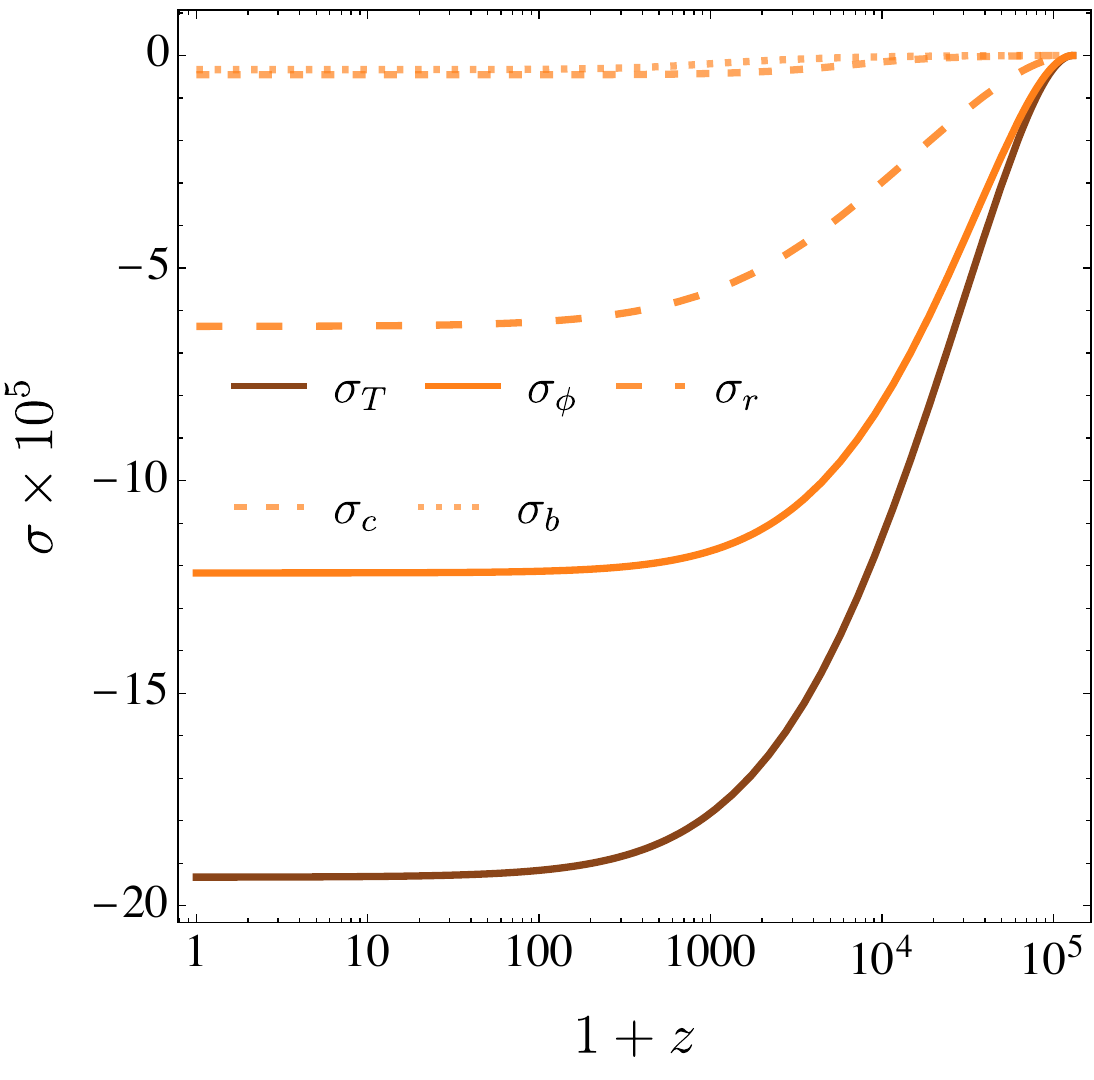}}
\caption{(Left) Contributions of each component to the total shear, $\Sigma_T$, during the expansion history. The contribution from dark energy ($\Sigma_\phi$) remains negligible, while the contributions from baryons ($\Sigma_b$) and dark matter ($\Sigma_c$) complement the dominant contribution from radiation. (Right) Contribution of each component to the total shear, $\sigma_T$, obtained upon integrating each contribution to $\Sigma_T$.}
\label{Fig: Shear IDE}
\end{figure*}

The left panel of Fig.~\ref{Fig: IDE Dynamics} presents the cosmological evolution of the density parameters $\Omega_r$, $\Omega_m \equiv \Omega_c + \Omega_b$, and $\Omega_\text{DE}$, as well as the equations of state $w_\text{eff}$ and $w_\text{DE}$. The results demonstrate that the known thermal history of the Universe remains unaltered by the non-comoving components. The right panel illustrates the evolution of the variables $|x_1|$ and $x_3$. At high redshifts, during the radiation-dominated epoch where $\Omega_r$ dominates and $w_\text{eff} \approx 1/3$, the hierarchy $x_3 \gg |x_1|$ is evident. During the matter-dominated era ($w_\text{eff} \approx 0$), $w_\text{DE}$ briefly crosses the phantom line. At late times, $|x_1|$ dominates the scalar field's energy content, leading the system to the de Sitter attractor, characterized by $w_\text{eff} = w_\text{DE} = -1$, marking the onset of exponential expansion.

The left panel of Fig.~\ref{Fig: Shear IDE} illustrates the contributions of various cosmic components to the total shear, $\Sigma_T$. Starting with $\Sigma_i = 0$, the relative motion of these components induces shear, initially dominated by the scalar field contribution, $\Sigma_\phi$. Meanwhile, smaller contributions from baryons and dark matter ($\Sigma_b$ and $\Sigma_c$, respectively) complement the more significant contribution from radiation ($\Sigma_r$). Despite this interplay, the total shear remains well-controlled and decays over time. At late times, $\Sigma_T$ approaches zero, corresponding to the isotropic attractor of the system. From the numerical analysis, the present-day shear is estimated to be $\Sigma_0 \approx 10^{-10}$, which is significantly below the observational upper limit of $|\Sigma_0| < 10^{-3}$, as derived from supernova constraints~\cite{Campanelli:2010zx, Amirhashchi:2018nxl}.

The right panel of Fig.~\ref{Fig: Shear IDE} shows the corresponding geometric shear, $\sigma$, obtained by integrating $\Sigma$. The analysis reveals that all components asymptote to a constant value, collectively determining $\sigma_T$. This constancy reflects the isotropization of the cosmic expansion, as such a constant can be absorbed into a redefinition of the spatial coordinates in the Bianchi I metric [Eq.~\eqref{Eq: Bianchi I}], effectively reducing it to the FLRW form. While the shear decays over time, the early anisotropy imprints itself on the Universe's evolution, influencing observational signatures such as the CMB quadrupole amplitude. Using Eq.~\eqref{Eq: CMB quadrupole}, the contribution of this particular realization to the CMB quadrupole amplitude is estimated as:
\begin{equation}
    \frac{(\delta T)_\text{quad}}{T_0} \lesssim |\sigma_\text{dec} - \sigma_0| \approx 1.6 \times 10^{-6}.
\end{equation}
This result aligns with the conservative observational bound for the CMB quadrupole amplitude, $\left| \sigma_\text{dec} - \sigma_0 \right| < 10^{-4}$~\cite{Appleby:2009za}. If the observed CMB dipole originates entirely from the bulk flow generated by the KGB motion, then the velocity relative to the CMB frame is $v_r = 1.23 \times 10^{-3}$. This motion leaves a residual imprint on the CMB quadrupole amplitude of the order of $10^{-6}$, consistent with observational constraints.

To conclude this section, we provide a numerical evaluation of the general estimates for an arbitrary KGB model derived in Sec.~\ref{Sec: quadrupole}. There, the radiation's contribution to anisotropy was estimated as:
$$
  (\Delta \sigma)_r \simeq -\frac{8}{3}\frac{\Omega_{r, 0} v_r^2}{a_{\text{dec}}} \approx -4.6 \times 10^{-7}.
$$
Integrating $\Sigma_r$ from the dynamical analysis for the imperfect dark energy scenario, we find $(\Delta \sigma)_r \approx -8.8 \times 10^{-7}$, showing good agreement with the analytical estimate. On the other hand, the scalar field contribution was estimated as: 
$$
(\Delta \sigma)_{\text{KGB}} \lesssim 2 \frac{\lambda \bar{Q}}{a_\text{dec}}\left\vert\frac{J^z}{\mc{J}^0}\right\vert_{\max} \approx 10^{-4} \left\vert\frac{J^z}{\mc{J}^0}\right\vert_{\max}.
$$
For the imperfect dark energy model, the numerical analysis yields:
\begin{equation}
    \left\vert\frac{J^z}{\mc{J}^0}\right\vert_{\max} \approx \frac{5}{6} \sqrt{u_{\max}} \approx \frac{5}{6} \left\vert \frac{\lambda}{\phi'}\right\vert_{\max} \approx 0.42,
\end{equation}
confirming the estimation $J^z / \mc{J}^0 \sim \lambda /\phi'$. Since the CMB quadrupole is $\sim 10^{-5}$, the moving KGB model can yield a non-negligible contribution to the quadrupole if an epoch exists where $J^z \simeq 10^{-1} \mc{J}^0$, consistent with our numerical findings.

Finally, for this specific KGB realization, we numerically estimate the constants derived in Eqs.~\eqref{Eq: Q estimation} and \eqref{Eq: lambda estimation}. Specifically, we find $Q \lesssim 2.3 \times 10^{-4} \, \mpl H_0$ and $\lambda \lesssim 3 \times 10^{-3} \, \mpl H_0$. Expressing $\lambda$ in terms of the dynamical variables:
\begin{equation} 
\lambda = \sqrt{-6 u_0 x_{1, 0}} \, \mpl H_0. 
\end{equation}
Using the numerical solution and assuming $\lambda \bar{Q} = 1.476 \times 10^{-7}$, we find:
\begin{equation} 
\lambda \approx 5.77 \times 10^{-4} \, \mpl H_0, \quad Q \approx 7.63 \times 10^{-4} \, \mpl H_0, 
\end{equation}
which are in good agreement with the analytical estimations. Similarly, the mass scale $\Lambda^3$ can be computed as:
\begin{equation}
    \Lambda^3 = \left( - 6 \frac{x_{1, 0}}{x_{3, 0}^{2/3}} \right)^{3/2} \, \mpl H_0^2 \approx 6.12 \, \mpl H_0^2,
\end{equation}
which is close to our estimation $\Lambda^3 \simeq 3\sqrt{6} \, \mpl H_0^2 \approx 7.35 \, \mpl H_0^2$.

\section{Conclusions}
\label{Sec:conclusions}

In this work, we have investigated cosmological scenarios based on shift-symmetric Horndeski theories with an inhomogeneous scalar field profile  $\langle \phi \rangle = \phi(t) + \lambda_i x^i$ that realises homogeneity as a combination of spatial translations and internal shifts. Although the configuration preserves homogeneity, the presence of the constant vector $\vec{\lambda}$ introduces a preferred direction so that the resulting cosmologies are described by an axisymmetric Bianchi I metric. The existing preferred direction is directly related to the direction of momentum density of the Horndeski field which, remarkably, follows the simple and elegant relation $T^0{}_i = -J^0\lambda_i$, where $J^0$ is the temporal component of  the conserved current associated to the shift symmetry. Since we have the universal relation $\sqrt{-g}J^0=Q$, with $Q$ the conserved charge, we have found that the momentum density has the universal evolution law $\sqrt{-g} \, T^0{}_i=-Q\lambda_i$ so it is fully determined by the charge $Q$ and the constant vector $\vec{\lambda}$.  We have shown that this relation follows from the off-shell Bianchi identities for homogeneous configurations of a generic class of shift-symmetric scalar-tensor theories (beyond the Horndeski realm). An analogous identity holds for theories with vector fields but with a crucial difference, namely: $J^0$ corresponds to the equation of motion of the vector field temporal component and, therefore, it vanishes on-shell. This property obstructs analogous solutions with vector fields and explains an intriguing feature that has been found in the literature for generic theories with vector fields.   

Since the homogeneous and anisotropic cosmological solutions supported by the considered inhomogeneous scalar field profile are characterised by a non-vanishing momentum density, these scenarios provide an explicit realisation of cosmologies with non-comoving components and, in particular, moving dark energy models. This is the reason that motivated us to dub the constructed solutions moving Horndeski models. An important property of these scenarios is that at least a second component in the universe is required for the non-triviality of the inhomogeneous piece of the scalar field. Thus, the moving Horndeski component describes the relative motion between the other components in the universe and the Horndeski field (or its homogeneous rest frame). We have constructed the CCM wherein a constraint between the velocities of all the components and the Horndeski field exists. We have analised the mini-superspace formulation of these scenarios, showing that the CCM rest frame amounts to trivialising the shift vector and, hence, it can only be chosen after obtaining the equations of motion, but it is crucial to keep it in the mini-superspace action to recover the momentum constraint. The dual formulation in terms of 2-form fields has also been constructed.

The moving DE paradigm generates a cosmological component for the CMB dipole, that is not fully generated by the motion with respect to the CMB but there is an additional contribution due to the relative motion with respect to the CCM, and for the quadrupole, due to the anisotropic expansion generated by the relative motions. Although higher order multipoles can also be affected, the smallness of the velocities make those contributions completely negligible. To explore these contributions in our moving Horndeski scenario, we have specialized our analysis to the KGB subclass of Horndeski theories, which aligns with current constraints from gravitational wave observations. Within this framework, we have calculated the contributions to the CMB dipole and quadrupole. Using the universal relation for the momentum density, we showed that the dipole contribution is determined by $\lambda Q$, making it independent of the specific realization of the KGB theory and providing a robust prediction. The quadrupole contribution, however, depends on the evolution of shear from decoupling to the present epoch, introducing model dependence. However, we have estimated the required condition to avoid constraints from the CMB quadrupole for the generic case. On the other hand, large-scale bulk flows also emerge within the framework of moving Horndeski models for dark energy as a consequence of the different rest frames of matter and radiation induced by the presence of the moving Horndeski field. Thus, the observational claims of large-scale bulk flows~\cite{Watkins:2008hf, Watkins:2023rll, Lopes:2024vfz} are naturally accommodated in the moving Horndeski models and make them compelling candidates for addressing such observations that could potentially challenge the standard model predictions. We have shown that these observations can be explained by the moving KGB scenarios while being compatible with the CMB dipole and quadrupole. In order to illustrate our general results, we have considered a specific realisation of moving KGB dark energy model that can give non-negligible contributions to the large scale bulk flows and the CMB dipole while avoiding quadrupole CMB constraints.

The broader implications of our work lie in the connection between the moving DE framework and observational evidence for anisotropies, bulk flows, dipolar modulations and preferred directions in the universe. These observations challenge the isotropy assumption of the standard cosmological model and motivate the exploration of scenarios beyond the standard isotropic realm like the one presented here. Our framework provides a field-theoretic underpinning for these phenomena, offering a compelling and predictive approach to studying anisotropic cosmologies.

In summary, this work establishes moving DE as a compelling framework for understanding large-scale bulk flows and anisotropic effects in the universe. By grounding these phenomena in the context of shift-symmetric Horndeski theories, we provide a robust theoretical foundation for interpreting cosmological observations and exploring new dynamics beyond the standard model of cosmology. The natural step that would follow would be a thorough analysis of the evolution of the cosmological perturbations. This can be done directly by considering perturbations around a Bianchi-I universe or as a double perturbative expansion in the velocities of the components and the inhomogeneous perturbations. In both cases, despite interesting observational signatures in relevant observables such as the growth of structures, the motion of the KGB field will also affect the stability conditions of these models. In this respect, the permitted parameter space will be non-trivially modified by the new parameter $\lambda$ that determines the motion of the KGB field. We leave these issues for future work. 

\section*{Acknowledgements}

We thank Antonio L. Maroto and Shinji Tsujikawa for useful discussions. B.O.Q. is supported by Vicerrector\'ia de Investigaciones - Universidad del Valle Grant No. 71373. J.B.J. was supported by the Project PID2021-122938NB-I00 funded by the Spanish “Ministerio de Ciencia e Innovaci\'on" and FEDER “A way of making Europe” and the Project SA097P24 funded by Junta de Castilla y Le\'on.

\section*{Appendices}
\appendix

\section{Moving Horndeski in the mini-superspace}
\label{App:MovingHorndeskiMinisuperspace}
In this Appendix, we report the explicit form of the minisuperspace action in Eq.~\eqref{Eq:minisuperspaceaction} that is given by:
\be
\mathcal{S}=\int\dd t a^3N\left[P(\mathcal{X})+\sum_{i=2}^5\mathcal{L}_i\right],
\ee
with:
\begin{align}
\mathcal{L}_2 =& \, G_2(X),\\
\mathcal{L}_3 =& \, \frac{G_3(X)}{N^2}\left[\dot{N}_\phi+\left(3\frac{\dot{a}}{a}-\frac{\dot{N}}{N}\right)N_\phi\right],\\
\mathcal{L}_4 =& \, \frac{6 G_4(X)}{N^2}\left[\frac{\ddot{a}}{a}+\frac{\dot{a}^2}{a^2}-\frac{\dot{a}\dot{N}}{aN}+\dot{\alpha}^2\right]
+\frac{6N_\phi G_{4X}(X)}{N^4}\left[\frac{\dot{a}}{a}\dot{N}_\phi+\left(\frac{\dot{a}^2}{a^2}-\frac{\dot{a}\dot{N}}{aN}-\dot{\alpha}^2\right)N_\phi\right]\nonumber\\
&+\frac{2\lambda^2 e^{4\sigma}G_{4X}(X)}{N^2 a^2}\left[\frac{\dot{a}^2}{a^2}-4\frac{\dot{a}}{a}\dot{\alpha}+4\dot{\alpha}^2\right],\\
\mathcal{L}_5 =& \, \frac{3G_5(X)}{N^4}\left[\left(\frac{\dot{a}^2}{a^2}-\dot{\alpha}^2\right)\dot{N}_\phi
+\left(\frac{\dot{a}^3}{a^3}-3\frac{\dot{a}^2\dot{N}}{a^2N}+2\frac{\dot{a}\ddot{a}}{a^2}+3\left(\frac{\dot{N}}{N}-\frac{\dot{a}}{a}\right)\dot{\alpha}^2-2\dot{\alpha}\ddot{\alpha}\right)N_\phi\right]\nonumber\\
&+\frac{N_\phi^2 G_{5X}(X)}{N^6}\left[3\left(\frac{\dot{a}^2}{a^2}-\dot{\alpha}^2\right)\dot{N}_\phi+\left(\frac{\dot{a}^3}{a^3}-3\frac{\dot{a}^2\dot{N}}{a^2N}+3\left(\frac{\dot{N}}{N}-\frac{\dot{a}}{a}\right)\dot{\alpha}^2-2\dot{\alpha}^3\right)N_\phi\right]\nonumber\\
&+\frac{2\lambda^2 e^{4\sigma}N_\phi G_{5X}(X)}{N^4 a^2}\left[\frac{\dot{a}^3}{a^3}-3\frac{\dot{a}^2}{a^2}\dot{\alpha}+4\dot{\alpha}^3\right],
\end{align}
where we have defined $N_\phi\equiv\dot{\phi}-\lambda N_z$. We can introduce the analogous variable for the super-fluid field $N_\chi\equiv\dot{\chi}-\mu N_z$ so we have
\be
X=\frac{N_\phi^2}{2N^2}-\frac{\lambda^2}{2a^2}e^{4\sigma},\quad \mathcal{X}=\frac{N_\chi^2}{2N^2}-\frac{\mu^2}{2a^2}e^{4\sigma}.
\ee
Thus, we corroborate the claim made below Eq.~\eqref{eq:Qconservationmini} that the derivatives of the scalar fields and the shift only enter through the combinations $N_\phi$ and $N_\chi$ so that we can relate the shift equation of motion with the conserved charges. As explained in Sec.~\ref{sec:mini-superspace}, this relation stems from a residual symmetry. The resulting mini-superspace action can be further simplified by performing integrations by parts to make more apparent the propagating degrees of freedom and the constraints, but we have not pursued such a task.

\section{Horndeski theory after GW170817}
\label{App: GWs in Horndeski}

In Sec.~\ref{Sec: Horndeski in motion}, we introduced the full Horndeski action represented in Eqs.~\eqref{Eq: L2}-\eqref{Eq: L5}. However, the discovery of gravitational waves (GWs) by the LIGO collaboration~\cite{LIGOScientific:2017vwq}, and the precise determination of their propagation speed~\cite{Liu:2020slm, Baker:2022eiz} have imposed stringent constraints on modifications to gravity at late cosmic times~\cite{Ezquiaga:2017ekz,Creminelli:2017sry, Baker:2017hug, Kreisch:2017uet}. 

The propagation speed of GWs is governed by the evolution equation for tensor modes. These tensor modes arise from perturbations in the metric, which in Cartesian coordinates takes the form:  
\begin{equation}
\label{Eq: Tensor modes}
\text{d} s^2 = a^2(\eta) \left[ - \text{d} \eta^2 + ( \delta_{i j} + h_{i j} )  \text{d} x^i \text{d} x^j \right],
\end{equation}
where $h_{i j}$ are the tensor perturbations. The evolution equation for these tensor modes can be written in general as: 
\begin{equation}
\label{Eq: GW Equation}
\( h^i{}_j \)'' + \( 2 \mc{H} + \gamma_T \) \( h^i{}_j \)' - c_T^2 \nabla^2 h^i{}_j = 0.
\end{equation}
This equation describes the propagation of GWs as waves with a propagation speed $c_T$, subject to the friction term $(2 \mc{H} + \gamma_T)$ due to the universe's expansionary dynamics.

Considering the full Lagrangian of the Horndeski theory in Eqs.~\eqref{Eq: L2}-\eqref{Eq: L5}, and allowing small inhomogeneities in the scalar field, we can derive the gravitational field equations for the tensor modes up to first order. The propagation speed of gravitational waves is then given by:
\begin{align}
c_T^2 &= \frac{2 a^4 G_{4} - a^2 \phi'^2 G_{5\phi} + \mathcal{H} \phi'^3 G_{5 X} - \phi'^2 \phi'' G_{5 X}}{2 a^4 G_{4} - 2 a^2 \phi'^2 G_{4 X} + a^2 \phi'^2 G_{5\phi} - \mathcal{H} \phi'^3 G_{5 X}},
\end{align}
which is consistent with the result found in Ref.~\cite{Kobayashi:2011nu, Cardona:2022lcz}. Observational data indicates that the propagation speed of gravitational waves is effectively the speed of light. To ensure that general Horndeski theories satisfy the condition $c_T^2 = 1$ for all times and scales without requiring fine-tuning,\footnote{In Bayesian statistics, models that require fine-tuning to fit observational data are penalized by the Occam factor~\cite{Mckay2003}.} the free functions in the general Lagrangian must satisfy the following condition:\footnote{Under the condition in Eq.~\eqref{Eq: GW constraint}, the drag term $\gamma_T$ vanishes, resulting in the standard wave equation for a massless field propagating at the speed of light.}
\begin{equation}
\label{Eq: GW constraint}
G_4 = G_4 (\phi), \qquad G_5 = \text{constant}.
\end{equation}
Since $G_5$ is constant and the Bianchi identiy holds ($\nabla_\mu G^{\mu\nu} = 0$), the term $\mc{L}_5^\text{ST}$ in the Lagrangian reduces to a total derivative:
\begin{equation}
    \mc{L}_5^\text{ST} = \nabla_\mu ( G_5 \, G^{\mu\nu} \, \nabla_\nu \phi),
\end{equation}
which does not contribute to the dynamics. Therefore, the remaining viable Horndeski theories are described by:
\begin{align}
\mc{L}_2^\text{ST} &= G_2 (\phi, X), \qquad \mc{L}_3^\text{ST} = - G_3 (\phi, X_1) \square \phi, \qquad \mc{L}_4^\text{ST} = G_4 (\phi) R. \label{Eq: L4}
\end{align} 
In the moving Horndeski framework introduced in Sec.~\ref{Sec: Horndeski in motion}, these terms are further simplified by assuming a shift symmetry of the scalar field $\phi \mapsto \phi + c$. Therefore, the free functions $G_2$, $G_3$ depend only on the kinetic term $X$ and $G_4$ is a constant. This shift-symmetry then leads to the the effective Lagrangian:
\begin{equation}
\label{Eq: Eff Lagrangian}
\mc{L} = \mc{L}_\text{EH} + \mc{L}_\text{KGB},
\end{equation}
where we have chosen the constant $2 G_4 = m_\text{P}^2$ in order to reproduce the Einstein-Hilbert Lagrangian $\mc{L}_\text{EH} \equiv m_\text{P}^2 R/2$ from the term $\mc{L}_4^\text{ST}$. The remaining scalar-tensor interactions are encoded in the so-called kinetic gravity braiding (KGB) model (or cubic Horndeski) defined here as $\mc{L}_\text{KGB} \equiv G_2(X) - G_3(X) \Box \phi$.

\section{Dynamical analysis}
\label{App: Dynamical System}

Here, we elaborate on the details about dynamical system analysis introduced in Sec.~\ref{Sec: Moving KGB}. 

The expansion dynamics are governed by the first Friedman equation. For the imperfect dark energy scenario defined by the Lagrangian in Eq.\eqref{Eq: Imperfect DE}, this equation is expressed as: 
\begin{equation}
    3 m_\text{P}^2 \mc{H}^2 = a^2\left(\rho_r + \rho_b + \rho_c + \rho_\phi \right) + 3 m_\text{P}^2 \sigma'^2,
\end{equation}
The dark energy density, $\rho_\phi$, is derived from the ``$00$'' component of the energy-momentum tensor $T_{\mu\nu}^{(\phi)}$ in Eq.~\eqref{Eq: KGB Energy Tensor}, yielding:
\begin{equation}
    \rho_\phi = -\frac{\phi'^2}{2 a^2} + \frac{3 \mc{H} \phi'^3}{\Lambda^3 a^4} - \frac{\lambda^2 e^{4\sigma}}{a^2} \left\{\frac{1}{2} + \frac{\mc{H}\phi'}{\Lambda^3 a^2} - \frac{2\sigma'\phi'}{\Lambda^3 a^2} \right\}.
\end{equation}
Dividing the Friedman equation by $3 m_\text{P}^2 \mc{H}^2$, the dynamical variables introduced in Eqs.~\eqref{Eq: Variables} are defined, allowing the Friedman equation to be reformulated as a constraint. The dark matter density parameter is expressed as $\Omega_c = 1 - \Omega_r - \Omega_b - \Omega_\phi$, with $\Omega_\phi$ given by:
\begin{equation}
    \Omega_\phi \equiv \frac{a^2 \rho_\phi}{3 m_\text{P}^2 \mc{H}^2} = x_1 + x_2 + x_3 + \frac{1}{3}u \left(3x_1 - 2x_2 - x_3 + 2 \Sigma x_3 \right).
\end{equation} 
Differentiating these variables with respect to conformal time and normalizing by $\mc{H}$ provides the system of first-order equations in Eqs.~\eqref{Eq: x_1 Eq}-\eqref{Eq: Sigma Eq}. To close the system, we introduce the parameters $\epsilon_\phi \equiv \phi''/(\mc{H} \phi')$ and $h_\phi \equiv \mc{H}'/\mc{H}^2$. The parameter $h_\phi$ is determined from the second Friedman equation:
\begin{equation}
    - 6 m_\text{P}^2 \mc{H}' = a^2 \left( 2\rho_r + \rho_b + \rho_c + \rho_\phi + 3 p_\phi \right) + 12 m_\text{P}^2 \sigma'^2,
\end{equation}
where $p_\phi$ is the dark energy pressure derived from the trace of $T_{\mu\nu}^{(\phi)}$ in Eq.~\eqref{Eq: KGB Energy Tensor}:
\begin{equation}
\label{Eq: IDE p}
    p_\phi = -\frac{\phi'^2}{2 a^2} + \frac{\phi'^2}{\Lambda^3 a^2} \left( \frac{\mc{H} \phi'}{a^2} - \frac{\phi''}{a^2} \right) + \frac{\lambda^2 e^{4\sigma}}{a^2} \left\{ \frac{1}{6} + \frac{1}{3\Lambda^3}\left( \frac{\phi''}{a^2} + \frac{\mc{H}\phi'}{a^2} - 6\frac{\sigma'\phi'}{a^2} \right) \right\}.
\end{equation}
From this, the dark energy equation of state, $w_\text{DE} \equiv p_\phi / \rho_\phi$, is expressed in terms of the dimensionless variables as in Eq.~\eqref{Eq: wDE}. The parameter $\epsilon_\phi$ is derived from the scalar field equation of motion:
\begin{equation}
    0 = \frac{\phi''}{a^2} + \frac{2 \mc{H} \phi'}{a^2} - \frac{3\mc{H}' \phi'^2}{\Lambda^3 a^4}  - \frac{6 \mc{H} \phi' \phi''}{\Lambda^3 a^4} + \frac{\lambda^2 e^{4\sigma}}{\Lambda^3a^4}\left(\mc{H}' + 4 \mc{H}\sigma' - 8 \sigma'^2 - 2 \sigma'' \right).
\end{equation}

The system of equations presented in Eqs.~\eqref{Eq: x_1 Eq}-\eqref{Eq: Sigma Eq} is inherently non-autonomous, unlike typical cosmological systems~\cite{Bahamonde:2017ize}, because the velocities of pressureless matter components decay proportionally to $1/a$ following their decoupling from the primordial plasma. While the system’s autonomous nature could be restored by including evolution equations for these velocities, this would require expanding the phase space, making the analysis more complex. Instead, we focus on analyzing the system at very late times. 

At sufficiently late times, the energy densities of dust and radiation diminish significantly due to their dilution with the Universe's expansion. As a result, these components no longer play a substantial role in the anisotropic expansion dynamics. By neglecting their contributions in the evolution equation for $\Sigma$, the system can be effectively reformulated as an autonomous one. Under the assumptions $\Omega_r = \Omega_b = \Omega_c = 0$, the fixed points of the system correspond to stationary solutions where $x_1' = 0$, $x_3' = 0$, $u' = 0$, and $\Sigma' = 0$ in Eqs.~\eqref{Eq: x_1 Eq}-\eqref{Eq: Sigma Eq}. Solving these algebraic equations for the dynamical variables yields a unique fixed point relevant to the late-time accelerated expansion phase:
\begin{equation}
\label{Eq: dS Attractor}
    x_1 = - 1, \quad x_3 = 2, \quad u = 0, \quad \Sigma = 0, \quad w_\text{DE} = w_\text{eff} = -1, \quad \epsilon_\phi = h_\phi = 1.
\end{equation}
This fixed point represents an isotropic de Sitter solution, characterized by an exponential expansion of the Universe dominated by the scalar field’s energy density. Substituting the variables $x_1$ and $x_3$ in terms of the field $\phi'$, we derive a relation between the field and the mass scale:
\begin{equation}
    \phi' = \frac{a^2 \Lambda^3}{3 \mc{H}},
\end{equation}
which agrees with the expression in Eq.~\eqref{eq:dotphiattractor}, confirming that our leading-order approximation in $\lambda$ holds in general.

To integrate the non-autonomous system, we assume initial conditions deep in the radiation-dominated epoch [see Eqs.~\eqref{Eq: Init Cond}]. These initial conditions are chosen to align with the observed cosmic budget today. Additionally, the CMB dipole constraint in Eq.~\eqref{eq:dipolebound} imposes a condition on the constant $\lambda \bar{Q}$ that is saturated when $v_r = 1.23 \times 10^{-3}$. This constraint also affects the initial condition for $u$, since $u_i \approx \lambda / \phi_i'$ and $\phi_i'$ is determined by the variables $x_{1, i}$ and $x_{3, i}$. Therefore, the initial condition for $u$ is fixed rather than freely chosen. 

To compute $u_i$, we rewrite the constant $\lambda \bar{Q}$ in terms of the dynamical variables:
\begin{equation}
\label{Eq: lambda Q in Vars}
    \lambda \bar{Q} = \frac{1}{3}\sqrt{u} a^4 \frac{H^2}{H_0^2} \left\{ 6x_1 + 3x_3 - u x_3(1 - 2\Sigma) \right\}.
\end{equation}
During the radiation-dominated epoch, assuming $H^2 = H_0^2 \Omega_{r, 0} a^{-4}$ and substituting the initial conditions from Eq.~\eqref{Eq: Init Cond}, we estimate $u_i \approx 0.29$. After numerically solving the autonomous system with this value, we refine our solution by adjusting $u_i$ to satisfy $\lambda \bar{Q} = 1.467 \times 10^{-7}$. This iterative process yields a final value of $u_i = 0.26$.

\section{Another example: Galileon Ghost Condensate}
\label{App: GGC}

In this appendix, we examine the predictions of an alternative KGB model within the moving dark energy framework, specifically the Galileon Ghost Condensate (GGC) model. This model has been shown to achieve remarkable concordance with observational data~\cite{Peirone:2019aua}. However, as we demonstrate below, it fails to induce significant effects on the lowest CMB multipoles due to the scalar field's dynamics.

The GGC model is governed by the following Lagrangian:
\begin{equation} 
\mathcal{L}_\text{KGB} = -X + a_2 X^2 + \frac{X}{\Lambda^3} \Box \phi, 
\end{equation}
where $a_2$ is a dimensionful constant.  Due to its structural similarity to the imperfect dark energy scenario—differing only by the presence of the quadratic term in $X$—we adopt the dynamical system framework developed in Sec.~\ref{Sec:IDE} and Appendix~\ref{App: Dynamical System}, with minor modifications. Specifically, the relation for the constant $\lambda \bar{Q}$ in terms of the variables defined in Eq.~\eqref{Eq: lambda Q in Vars} is adjusted to:
\begin{equation}
\label{Eq: lQ for GGC}
    \lambda \bar{Q} = \frac{1}{3}\sqrt{u} a^4 \frac{H^2}{H_0^2} \left\{ 6x_1 + 4x_2 + 3x_3 - u \[4x_2 + x_3\(1 - 2\Sigma\)\] \right\}.
\end{equation}
During the radiation-dominated epoch, the Hubble parameter satisfies $H^2 = H_0^2 \Omega_{r, 0} a^{-4}$, and $\lambda \bar{Q}$ can be expressed in terms of the initial conditions of the variables and $\Omega_{r, 0}$. From Fig.1 in Ref.~\cite{Kase:2018iwp}, a set of initial conditions consistent with the observed cosmic budget at present is given by:
\begin{equation}
    x_{1, i} = -10^{-16}, \quad x_{2, i} = 3 \times 10^{-16}, \quad x_{3, i} = 10^{-9}, \quad \Omega_{r, i} = 0.975, \quad \Sigma_i = 0,
\end{equation}
at $z = 1.3 \times 10^{5}$. The value of $u_i$ is then fixed through $\lambda \bar{Q}$, which quantifies the contribution of the KGB field to the CMB dipole via Eq.~\eqref{eq:dipoleQlambda}. 

\begin{figure*}[t!]
\centering    
{\includegraphics[width=0.47\textwidth]{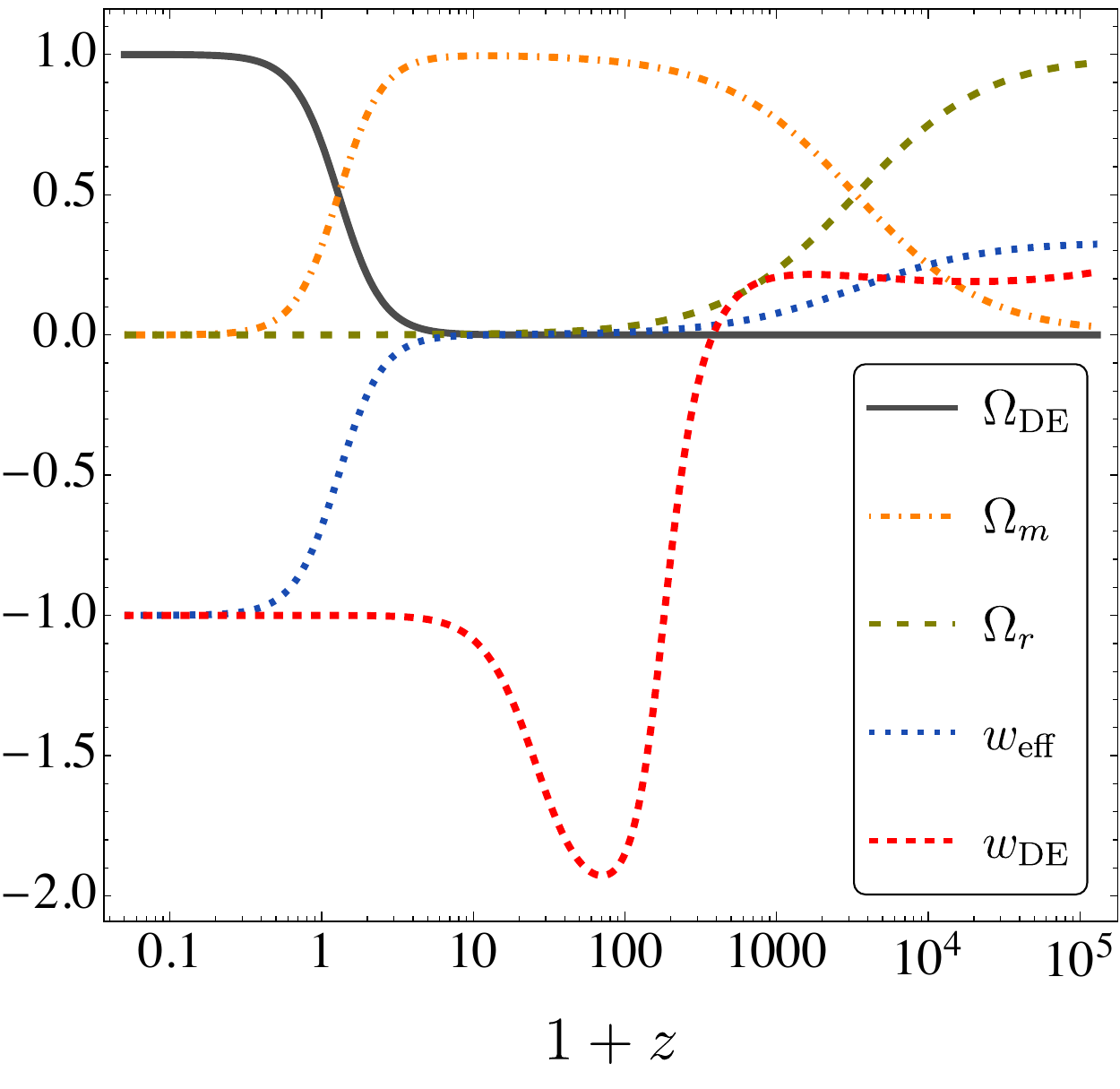}} \hfill
{\includegraphics[width=0.49\textwidth]{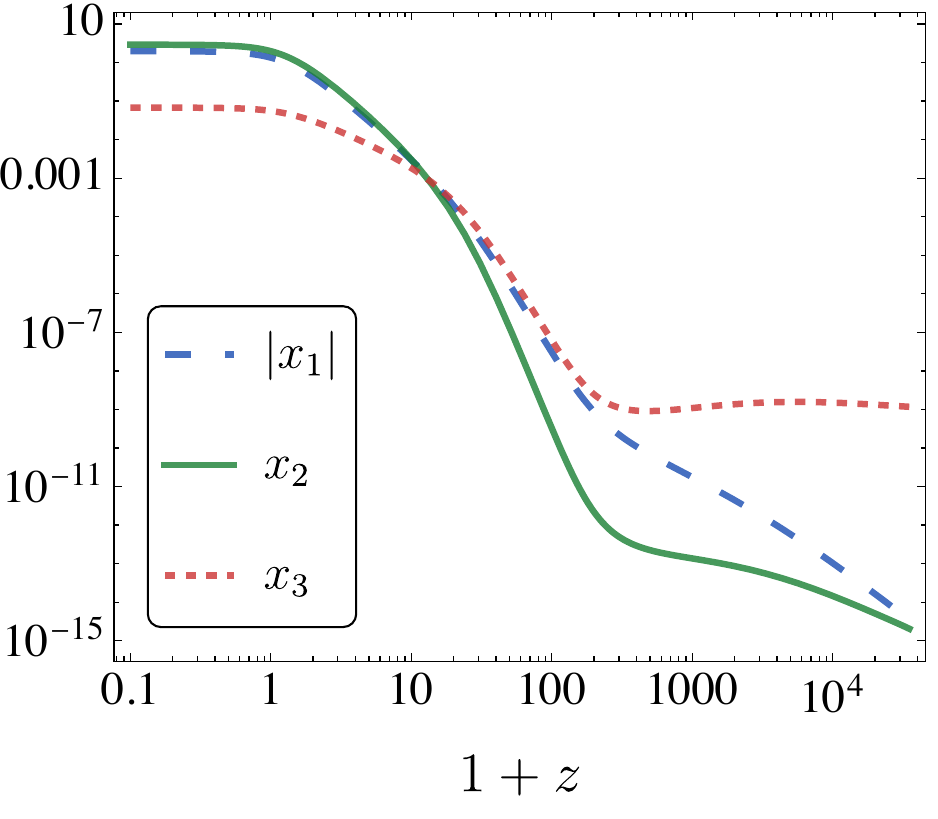}}
\caption{The time evolution of $\Omega_r$, $\Omega_m$, $\Omega_\text{DE}$, the equations of state $w_\text{eff}$ and $w_\text{DE}$ (in the left panel), and the variables $|x_1|$, $x_2$, and $x_3$ (in the right panel). At early times, when radiation is the dominant fluid in the cosmic budget ($w_\text{eff} \approx 1/3$), the scalar field evolution is dominated by $x_3$. At late times, $|x_1|$ and $x_2$ take over the dynamics, and the system transitions to a de Sitter expansion driven by the scalar field, with $w_\text{eff} = w_\text{DE} = -1$.}
\label{Fig: GGC Dynamics}
\end{figure*}

We found in Sec.~\ref{Sec:Dipole} that if the CMB dipole amplitude arises entirely from the KGB field, then $\lambda \bar{Q} \simeq 10^{-7}$ is required. However, substituting this value into Eq.\eqref{Eq: lQ for GGC} reveals no real solution for $u_i$. Alternatively, by assuming $u_i = 1$ (i.e., $\lambda = \phi'_i$), the initial conditions yield $\lambda \bar{Q} \simeq 10^{-14}$. Since $\lambda \bar{Q}$ determines the velocity of the radiation fluid via Eq.\eqref{Eq: Velocity radiation}, this implies $v_r \simeq 5 \times 10^{-10}$. Given the observed CMB dipole amplitude, $(\delta T)_\text{dip} / T_0 = 1.23 \times 10^{-3}$, the contribution from the KGB field is entirely negligible. On the other hand, for the CMB quadrupole, the contribution from the radiation fluid is proportional to $v_r^2$, yielding a value of order $10^{-20}$, which is also negligible. The KGB field's contribution to the quadrupole is constrained by Eq.~\eqref{Eq: sigma KGB bound}:  
\begin{equation}
    (\Delta \sigma)_\text{KGB} \lesssim \frac{\lambda \bar{Q}}{a_\text{dec}} \left\vert \frac{J^z}{\mathcal{J}^0}\right\vert_\text{max} \approx 10^{-11} \left\vert \frac{J^z}{\mathcal{J}^0}\right\vert_\text{max}.
\end{equation}  
Achieving $(\Delta \sigma)_\text{KGB} \sim 10^{-5}$ would require $J^z \sim 10^6 \mathcal{J}^0$ at some point during cosmic evolution, which is implausible within the GGC model. 

To confirm that the selected initial conditions result in an accurate expansion history, we illustrate the cosmological evolution of the density parameters $\Omega_r$, $\Omega_m$, and $\Omega_\text{DE}$, as well as the effective and dark energy equations of state, $w_\text{eff}$ and $w_\text{DE}$, in the left panel of Fig.~\ref{Fig: GGC Dynamics}. To account for the role of the variable $x_2$, we include its influence on the parameters $\epsilon_\phi$ and $h_\phi$, which modifies the autonomous system in Eqs.~\eqref{Eq: x_1 Eq}-\eqref{Eq: Sigma Eq} as follows:
\begin{align}
    \frac{\dd x_2}{\dd N} &= - 2 x_2 \( 1 + h_\phi - 2\epsilon_\phi \), \\
    \frac{\text{d} \Sigma}{\text{d}N} &= -\Sigma(2 + h_\phi) - \frac{1}{3}u\left[6x_1 + 4x_2 + x_3(2 + \epsilon_\phi)\right] - \sum_{\alpha = b, c, r} \(1 + w_{(\alpha)}\) \Omega_{(\alpha)} v^2_{(\alpha)}.
\end{align}
The right panel displays the evolution of the scalar field variables $|x_1|$, $x_2$, and $x_3$. At high redshifts, when the Universe is radiation-dominated, the hierarchy $x_3 \gg \{|x_1|, x_2\}$ is evident, leading to $\vert J^z/\mathcal{J}^0 \vert \propto \sqrt{u} \sim \vert \lambda/\phi' \vert \approx 1$. At late times, the dynamics of the scalar field are governed by $|x_1|$ and $x_2$, and the system evolves toward a de Sitter attractor where $w_\text{eff} = w_\text{DE} = -1$.

We can thus conclude that the cosmology of the Galileon ghost condensate does not provide a moving KGB model with observational signatures because the scalar field contributes negligibly to the energy density budget around decoupling time. 

\bibliographystyle{JHEPmodplain}
\bibliography{paper.bib}

\end{document}